\documentclass[a4paper,11pt]{article}
\pdfoutput=1
\usepackage{jheppub}

\usepackage{multirow, graphicx,amssymb,url,mathrsfs,amsmath}
\usepackage{wrapfig,boxedminipage,setspace,subfigure,epsfig}
\usepackage{amsxtra,amstext,latexsym,dsfont,amsfonts}
\usepackage{color,eucal}

\newcommand{\dd}{\mathrm{d}}

\newcommand{\grandpot}{{G}}
\newcommand{\tQ}{{\tilde{Q}}}
\newcommand{\tmu}{{\tilde{\mu}}}
\newcommand{\tbeta}{{\tilde{\beta}}}
\newcommand{\ttt}{{\tilde{t}}}
\newcommand{\trr}{{\tilde{r}}}
\newcommand{\txx}{{\tilde{x}}}
\newcommand{\tyy}{{\tilde{y}}}
\newcommand{\tTemp}{{\tilde{T}}}
\newcommand{\tss}{{\tilde{s}}}
\newcommand{\trho}{{\tilde{\rho}}}

\title{\boldmath Diffusion and Butterfly Velocity at Finite Density}

\author{Keun-Young Kim}
\author{and Chao Niu}

\emailAdd{fortoe@gist.ac.kr}
\emailAdd{chaoniu09@gmail.com}

\affiliation{ School of Physics and Chemistry, Gwangju Institute of Science and Technology,
	Gwangju 61005, Korea
}

\abstract{We study diffusion and butterfly velocity ($v_B$) in two holographic models, linear axion and axion-dilaton model,   with a momentum relaxation parameter ($\beta$) at finite density or chemical potential ($\mu$). Axion-dilaton model is particularly interesting since it shows linear-$T$-resistivity, which may have something to do with the universal bound of diffusion. 
At finite density, there are two diffusion constants $D_\pm$ describing the coupled diffusion of charge and energy.   
By computing $D_\pm$ exactly, we find that in the incoherent regime ($\beta/T \gg 1,\ \beta/\mu \gg 1$) $D_+$ is identified with the charge diffusion constant ($D_c$) and $D_-$ is identified with the energy diffusion constant ($D_e$). 
In the coherent regime, at very small density, $D_\pm$ are `maximally' mixed in the sense that $D_+(D_-)$ is identified with $D_e(D_c)$, which is opposite to the case in the incoherent regime.
In the incoherent regime $D_e \sim C_- \hbar v_B^2 / k_B T$  where $C_- = 1/2$ or 1 so it is universal independently of $\beta$ and $\mu$. However, $D_c \sim C_+ \hbar v_B^2 / k_B T$ where $C_+ = 1$ or $ \beta^2/16\pi^2 T^2$ so, in general, $C_+$ may not saturate to the lower bound in the incoherent regime, which suggests that the characteristic velocity for charge diffusion may not be the butterfly velocity.  We find that the finite density does not affect the diffusion property at zero density in the incoherent regime.  }

\begin{document}
\maketitle
\flushbottom
\section{Introduction}

Strongly correlated systems display exotic properties compared to weekly correlated ones. More interestingly, some of exotic properties appear in various materials  with a remarkable degree of universality~\cite{Hartnoll:2016apf}. For example, in diverse strange metals (cuprates, pnictides, heavy fermions etc.), resistivity ($\rho$) is linear in temperature ($T$)
\begin{equation}
\rho \sim T \,,
\end{equation}
unlike ordinary metals where it is quadratic in temperature as explained by the Fermi liquid theory. In many high temperature superconductors, an empirical universal property so called Homes' law have been observed~\cite{Homes:2004wv, Zaanen:2004aa}. It is a universal relation between critical temperature ($T_c$), DC conductivity near the critical temperature ($\sigma_{\mathrm{DC}}(T_c)$), and superfluid density at zero temperature $\rho_s(T=0)$:
\begin{equation}
\rho_s(T=0) =  C \sigma_{\mathrm{DC}}(T_c) T_c\,,
\end{equation}
where $C$ is a material independent universal number.

While such interesting phenomena in strongly correlated systems are not easy to analyze theoretically, 
gauge/gravity duality or holographic methods~\cite{Zaanen:2015oix, Ammon:2015wua,Hartnoll:2016apf}  have been providing new and effective ways to study them. For linear-$T$-resistivity, there have been a lot of works and we refer to \cite{Hartnoll:2016apf}. For the Homes' law see \cite{Erdmenger:2015qqa,Kim:2015dna,Kim:2016hzi,Kim:2016jjk}. 
In this paper, we investigate another universal property, universal bounds of charge and energy diffusion constants of strongly correlated systems from perspective of holographic methods. 

It is interesting that these universal diffusion bounds may have something to do with aforementioned linear-$T$-resistivity and the Homes' law  via a fundamental universal relaxation timescale ($\tau_P$) so called `Planckian' time scale~\cite{Sachdev:2011cs, Zaanen:2004aa}  
\begin{equation} \label{Ptau}
\tau_P \sim \frac{\hbar}{k_B T} \,.
\end{equation}
A basic idea is as follows. Charge diffusion constant ($D_c$) is proposed to be related to a relaxation time ($\tau_P$) \cite{Hartnoll:2014lpa}:
\begin{equation} \label{DD1}
D_c  \gtrsim v^2 \tau_P \gtrsim v^2\frac{\hbar}{k_B T} \,,
\end{equation}
with a characteristic velocity scale.  Transport in{ \it incoherent} metals, where momentum is relaxed quickly, may be governed by diffusive physics of charge and energy rather than momentum.  To make a connection between transport and diffusion one can use the Einstein relation $\sigma = D_c \chi$, where $\chi$ is the charge susceptibility. The Einstein relation with \eqref{DD1} yields
\begin{equation} \label{nice1}
\sigma \gtrsim \chi v^2\frac{\hbar}{k_B T} \,.
\end{equation}
 If $D_c$ saturates to the bound and $\chi v^2$ is temperature independent, $\sigma \sim 1/T$ hence linear-$T$-resistivity is explained. Once linear-$T$-resistivity is satisfied, the Homes' law is reduced to $\rho_s(T=0) = \tilde{C} \chi v^2 $ where $\tilde{C}$ is a universal constant.\footnote {This is a version of the Tanner's law \cite{Zaanen:2004aa} with the identification $n(T_c) = \chi v^2$, where $n(T_c)$ is the charge density at the critical temperature. } Even though the universality of diffusion bounds itself is important it becomes more appealing because of the relation to other universalities such as linear-$T$-resistivity and the Homes' law.

At finite density, there are two diffusion constants $D_\pm$ describing the coupled diffusion of charge and energy
so the Einstein relation need to be generalized~\cite{Hartnoll:2014lpa} as follows\footnote{The conductivities may be diagonalized as in \cite{Davison:2015bea}.}.
\begin{eqnarray} 
D_+ D_- & = & \frac{\sigma}{\chi} \frac{\kappa}{c_\rho}   \,,  \label{DD3} \\
D_+ + D_- & = & \frac{\sigma}{\chi} + \frac{\kappa}{c_\rho}  \label{DD4}
+ \frac{T(\zeta \sigma - \chi \alpha)^2}{c_\rho \chi^2 \sigma} \,,
\end{eqnarray}
where  $\sigma, \alpha, \kappa$ are the electric, thermoelectric and thermal conductivity respectively. 
$\chi$ is the compressibility, $c_\rho$ is the specific heat at fixed charge density and $\zeta$ is the thermoelectric susceptibility. If the charge density is zero, since $\alpha=\zeta=0$, $D_\pm$ are decoupled and $D_+$ and $D_-$ can be identified with the charge diffusion constant ($D_c$)and the energy diffusion constant ($D_e$) respectively.  

In \cite{Hartnoll:2014lpa}, it was proposed the diffusion constants are bounded as
\begin{equation} \label{DD2}
D_\pm \gtrsim v^2\frac{\hbar}{k_B T} \,,
\end{equation}
where $v$ is an unknown characteristic velocity.  
It was conjectured by noticing that the KSS(Kovtun, Son and Starinets) bound of shear viscosity per entropy ratio ($\eta/s$) at zero chemical potential in a relativistic system may be expressed as $D \gtrsim c^2\frac{\hbar}{k_B T} $, where $c$ is the speed of light and $D$ is the momentum diffusion constant.
\eqref{DD2} suggests that diffusion is governed by the Planckian time scale \eqref{Ptau} independently of the charge density and the mechanism of momentum relaxation.  Recently, this time scale has been observed in the scattering rates of materials showing a linear $T$ resistivity~\cite{Bruin804} and in the thermal diffusivity~\cite{Zhang:2016aa}. 

To investigate this conjectured bounds further, we first need to identify what the characteristic velocity ($v$) is in \eqref{DD2}. An interesting candidate is the butterfly velocity ($v_B$), the speed at which the chaos spatially propagates through the system~\cite{Blake:2016wvh}. It implies that there is some connection between transport properties at strong coupling and quantum chaos\footnote{While this connection between transport properties and chaos was first proposed in the holographic models, it has been also observed in condensed matter systems \cite{Aleiner:2016aa, Swingle:2016aa, Patel:2016aa, Zhang:2016aa}.}.
 This idea was first tested at zero density in a class of holographic model with an scaling infrared geometry in \cite{Blake:2016wvh, Blake:2016sud}, where  concrete examples supporting the bound \eqref{DD2} with the butterfly velocity were provided. More evidence for the energy diffusion bound ($D_e/v_B^2$) was shown in holographic models that flow to AdS$_2 \times R^d$ fixed points in the infrared in \cite{Blake:2016jnn} and in the Sachdev-Ye-Kitaev (SYK) models \cite{Gu:2016oyy, Davison:2016ngz, Jian:2017unn}\footnote{In \cite{Jian:2017unn}, it was also shown that the diffusion constant may vanish across the phase transition, implying a dynamical transition to an many-body localization phase.}. However, it was shown that charge diffusion ($D_c/v_B^2$) may not have a universal lower bound in 
striped holographic matter \cite{Lucas:2016yfl} and in the SYK model~\cite{Davison:2016ngz}. When the higher derivative 
correction is added the energy diffusion ($D_e/v_B^2$) still can have a lower bound while  the charge diffusion ($D_c/v_B^2$) may vanish depending on the higher derivative couplings~ \cite{Baggioli:2016pia}. However, recently, it was shown that the energy diffusion ($D_e/v_B^2$) also may not have a universal lower bound in 
an inhomogeneous SYK model~\cite{Gu:2017ohj}.

In this paper, our goal is to study the bounds \eqref{DD2} at {\it finite density}.  Most studies so far have focused on the case i) at zero density or  ii) $D_c$ and $D_e$ at finite density instead of $D_\pm$ by ignoring the mixing term, the third term in \eqref{DD4}, and/or by taking small temperature limit.   Unlike the previous studies, we first study $D_\pm$ at finite density {\it without any approximation} and deduce the property of $D_c$ and $D_e$ in the incoherent regime.  We consider two holographic models: the linear axion model~\cite{Andrade:2013gsa} and one of the axion-dilaton models~\cite{Gouteraux:2014hca,Caldarelli:2016nni} based on the Gubser-Rocha model~\cite{Gubser:2009qt}. We choose these models because both allow the analytic solutions and they are related in the sense that at zero density the axion-dilaton model undergoes the phase transition to the linear axion model if the momentum relaxation is weak. The axion-dilaton model is particularly interesting because this model exhibits linear-$T$-resistivity: it will be interesting to see if there is any relation between the universal bound of the charge diffusion and linear-$T$-resistivity such as \eqref{nice1}. 

This paper is organized as follows. In section \ref{sec2}, we summarize the methods and formulas we will use to compute the diffusion constants and the butterfly velocity.  In section \ref{sec3}, we study the linear axion model at finite density, focusing on i) the relation between $D_{\pm}$ and $D_c(D_e)$ and ii) the effect of finite density to diffusion and the butterfly velocity. 
In section \ref{sec4}, we first analyze the phase structure of a axion-dilaton theory based on the Gubser-Rocha model. We find that there are two branches of classical solutions. 
After figuring out the ground state, we study the diffusion constants and butterfly velocity both at zero and finite density.  In section \ref{conc}, we conclude.

\section{Methods} \label{sec2}
In this section, we briefly summarize the method and formulas we will use in our computation in section \ref{sec3} and \ref{sec4}. Our goal is to study the universal lower bound \eqref{DD2}, which can be written as
\begin{equation}
D_\pm  = \frac{C_\pm}{2\pi} v_B^2\frac{1}{ T} \,,
\end{equation}
where $C_\pm$ is expected to be universal and $2\pi$ is introduced for later convenience. From here, we set $\hbar=k_B=1$. In other words, our main objects are  
\begin{equation}
C_\pm = \frac{2\pi T D_\pm}{v_B^2} \,, \qquad 2\pi T D_\pm \  \mathrm{or}\  v_B^2 \,,
\end{equation}
which will be computed and displayed in section \ref{sec3} and \ref{sec4}. 

In this paper, we are mainly interested in the incoherent regime, $\beta/T \gg 1$ and $\beta/\mu \gg 1$, where
$\beta$ is the strength of momentum relaxation, $T$ is temperature and $\mu$ is chemical potential, because 
in this regime momentum is relaxed quickly and we expect the transport is governed by diffusion of charge and energy~\cite{Hartnoll:2014lpa}.

\subsection{Diffusion constants}
From \eqref{DD3} and \eqref{DD4} two diffusion constants are computed as
\begin{equation} \label{DDD0}
D_\pm = \frac{c_2 \pm \sqrt{c_2^2 - 4c_1}}{2} \,,
\end{equation}
where
\begin{equation}  \label{DDD1}
%\begin{split}
c_1  \equiv  \frac{\sigma}{\chi} \frac{\kappa}{c_\rho}    \,,   \qquad
c_2  \equiv  \frac{\sigma}{\chi} + \frac{\kappa}{c_\rho} + \mathcal{M} \,, \qquad 
\mathcal{M}\equiv \frac{T(\zeta \sigma - \chi \alpha)^2}{c_\rho \chi^2 \sigma} \,.
%\end{split}
\end{equation}
Six variables defining $c_1$ and $c_2$ belong to two classes: thermodynamic susceptibilities ($\chi, \zeta, c_\rho$) and conductivities ($\sigma, \kappa, \alpha$). 

First, thermodynamic susceptibilities are defined as 
\begin{equation} \label{sus1}
\begin{split}
& \chi \equiv - \frac{\partial^2 \grandpot}{\partial \mu^2} =  \left( \frac{\partial \rho}{\partial \mu} \right)_{T} \,, \\   
& \zeta \equiv - \frac{\partial^2 \grandpot}{\partial T \, \partial \mu} = \left( \frac{\partial \rho}{\partial T} \right)_{\mu}=\left( \frac{\partial s}{\partial \mu} \right)_{T}  \,, \\
& c_\rho \equiv c_\mu - \frac{\zeta^2 T}{\chi} \,, \qquad c_\mu \equiv - T \frac{\partial^2 \grandpot}{\partial T^2} = T \left( \frac{\partial s}{\partial T} \right)  \,,
\end{split}
\end{equation}
with the thermodynamic potential density at fixed chemical potential and temperature:
\begin{equation}
\grandpot = \epsilon - s T - \mu \rho.
\end{equation}
Once the thermodynamic potential is computed by the gravity on-shell action according to the AdS/CFT duality, the susceptibilities are computed by \eqref{sus1} following standard thermodynamics. 

Next, the DC conductivities of a class of holographic models with momentum relaxation  may be expressed  in terms of black hole horizon data~\cite{Donos:2014cya}. For the action
\begin{equation} \label{genact}
S=\int \mathrm{d}^4x\sqrt{-g} \left[R-\frac{Z(\phi)}{4} F^2 -\frac{1}{2}(\partial{\phi})^2-V(\phi) -\frac{1}{2}\sum_{I=1}^{2}(\Phi_I(\phi)\partial \psi_{I})^2  \right] \,,
\end{equation}
with the ansatz 
\begin{equation}\label{ans1}
\begin{split}
\mathrm{d} s^2 &= -  U \mathrm{d} t^2 +  \frac{\mathrm{d} r^2}{U}  + V_1\mathrm{d} x^2 + V_2\mathrm{d} y^2\,, \\ 
A&= a\mathrm{d} t    \,, \qquad \psi_I  = \beta_I \delta_{Ii} x^i\,,
\end{split}
\end{equation}
the DC electric ($\sigma$),  thermal ($\overline \kappa$) , and thermoelectric ($\alpha$) conductivities along the $x$-direction can be
computed at the black hole horizon ($r_h$) as follows\footnote{These DC formulas have been confirmed by computing the optical conductivities and taking the zero frequency limit~\cite{Kim:2014bza,Kim:2015sma,Kim:2015wba}.  See also  \cite{Blake:2013bqa, Blake:2013owa} which were the first papers developing the techniques to calculate the electric conductivity in terms of the black hole horizon data in massive gravity.}. 
\begin{equation} \label{gencond}
\begin{split}
\sigma&=\left[\frac{Z(\phi)s}{4\pi V_1}+\frac{4\pi\rho^2}{\beta_1^2\Phi_1(\phi)s}\right]_{r=r_h}\,,\\
\overline \kappa&=\left[\frac{4\pi sT}{\beta_1^2\Phi_1(\phi)}\right]_{r=r_h}\,, \\
\alpha&=\left[\frac{4\pi\rho}{\beta_1^2\Phi_1(\phi)}\right]_{r=r_h}\,,
\end{split}
\end{equation}
The thermal conductivity with open circuit boundary conditions, which is the usual thermal conductivity, is
\begin{equation}
\kappa = \overline \kappa - \frac{\alpha^2 T}{\sigma} \,.
\end{equation}

\subsection{Butterfly velocity}

In this section we briefly review on the butterfly velocity in strongly correlated systems and its holographic dual. 
For more details, we refer to \cite{Sekino:2008he, Shenker:2013pqa, Roberts:2014isa, Maldacena:2015waa, Blake:2016wvh, Roberts:2016wdl, Blake:2016sud, Ling:2016ibq, Alishahiha:2016cjk}. 

The butterfly effect  as chaotic behaviour refers to the exponential growth of a small perturbation to a quantum system. It can be diagnosed by certain out-of-time-order (four-point) correlation function (OTOC) of two generic Hermitian operators $W(t,\vec{x})$ and $V(0,0)$, or the following average of the commutator squared:
\begin{equation}
C(t,\vec{x}) \equiv -\langle \left[ W(t,\vec{x}), V(0,0) \right]^2 \rangle_\beta \,,
\end{equation}
where $\beta = 1/T$ (the inverse temperature) and $\langle \cdots \rangle_\beta$ denotes thermal average. 
The function $C(t,\vec{x})$ quantifies the effect of a perturbation $V(0,0)$ on $W(t,\vec{x})$ and characterizes the strength of the butterfly effect at $\vec{x}$ at time $t$ induced by a perturbation at $t=\vec{x}=0$.  In general, $C(t,\vec{x})$ takes the following form
\begin{equation}
C(t,\vec{x}) = e^{\lambda_L (t-t_* - \frac{|\vec{x}|}{v_B})} + \cdots \,.
\end{equation}
Here, $\lambda_L$ is called a ``Lyapunov'' exponent following the classical chaos terminology. It measures the rate at which the system becomes scrambled and lose memory of its initial state.  $t_*$ is the scrambling time at which $C(t,0)$ becomes order one. For $\vec{x} \ne 0$ there is a spatial delay in scrambling characterized by $v_B$, so called the ``butterfly velocity". $C(t,\vec{x})$ grows to be order one at $t = t_* + \frac{|\vec{x}|}{v_B}$, which defines an effective light cone for chaos, a ``butterfly effect cone''.  Outside of the cone $C(t,\vec{x}) \ll 1$, even if the operators $V$ and $W$ are time-like separated with respect to the causal light cone $t=\vec{x}$. Inside the cone,  $C(t,\vec{x})$ grows quickly. Therefore, the butterfly velocity $v_B$ characterizes the speed at which the chaos spatially propagates through the system.  

It has been shown that the Lyapunov exponent is bounded by the temperature
\begin{equation} \label{L11}
\lambda_L \le 2\pi k_B T / \hbar =  2\pi/\tau_P \,,
\end{equation}
and saturates to the bound $2\pi/\tau_P$ for thermal systems that have a dual holographic black hole description of which near horizon geometry is described by Einstein gravity.  Notice that the Planckian time scale appears as a time scale of the growth of chaos in time. Therefore, it is quite appealing to use the butterfly velocity as a characteristic velocity in \eqref{DD2}.  With this identification, the bound of diffusion constant can be written only in terms of the characteristic parameters in quantum chaos
\begin{equation} %\label{DD1}
D_\pm  \gtrsim v_B^2 \tau_P \gtrsim 2\pi v_B^2/\lambda_L  \,.
\end{equation}

Like the Lyapunov exponent, the butterfly velocity also can be computed holographically by considering shockwave geometries. 
For systems that have dual holographic model with an infrared geometry 
\begin{equation}
\dd s^2_{d+2} = - U(r) \dd t^2 + \frac{\dd r^2}{U(r)} + V(r) \dd \vec{x}_d^2 \,.
\end{equation}
The Lyapunov exponent and the butterfly velocity are given by \cite{Blake:2016wvh, Blake:2016sud}
\begin{equation} \label{BV1}
\lambda_L = \frac{2\pi }{\tau_P}\,, \qquad  v_B^2 = \frac{2\pi T}{V'(r_h)} \,,
\end{equation}
where $r_h$ is the location of horizon.  While the Lyapunov exponent is universal, saturating the bound \eqref{L11}, the butterfly velocity is not.

\newpage

\section{Linear axion model} \label{sec3}

We first consider a simple holographic model of momentum relaxation, the four dimensional linear scalar model~\cite{Andrade:2013gsa}. This is the Einstein-Maxwell action coupled to massless scalars: 
\begin{eqnarray}
S&=&\int
\mathrm{d}^4x\sqrt{-g} \left[ R+\frac{6}{L^2}-\frac{1}{4}F^2 -\frac{1}{2} \sum_{I=1,2} (\partial \psi_I)^2 \right] \,,  \label{eq:action}
\end{eqnarray}
where we have chosen a unit such that the gravitational constant $16 \pi G$ is equal to $1$.  The second term is nothing but $-2 \Lambda$ with the negative cosmological constant, $\Lambda = -\frac{3}{ L^2}$.  $F= \dd A$ is the field strength for a $U(1)$ gauge field $A$. Because we want to consider a system at finite density we assume $ A = A_t(r) \dd t $ as a background solution. In the last term, two massless scalar fields of the form $\psi_I =  \beta_{Ii} x^i = \beta \delta_{Ii} x^i$ are introduced to break the translational symmetry in an isotropic way in the $x$-$y$ space, where a constant $\beta$ is interpreted as the strength of momentum relaxation.   This term induces momentum relaxation effect so makes conductivity finite~\cite{Andrade:2013gsa}.  An advantage of this  ansatz for massless scalars is that we can still have a homogeneous metric background solution, even though the translational symmetry is broken.  Under these assumption for $A$ and $\psi_I$, 
a classical solution of the action \eqref{eq:action} is
\begin{equation}\label{axion}
\begin{split}
\mathrm{d} s^2 &= -  f(r) \mathrm{d} t^2 +  \frac{\mathrm{d} r^2}{f(r)}  + \frac{r^2}{L^2}(\mathrm{d} x^2 + \mathrm{d} y^2)\,, \\ 
& \quad f(r) = \frac{r^2}{L^2}\left(  1 + \frac{L^2\mu^2r_h^2 }{4r^4} -\frac{L^4\beta^2 }{2r^2}  -\frac{r_h^3}{r^3} \left( 1 + \frac{L^2\mu^2 }{4r_h^2} -\frac{L^4\beta^2 }{2r_h^2}  \right)   \right)  \,, \\
A&= \mu \left(  1- \frac{r_h}{r}   \right)\mathrm{d} t    \,, \qquad \psi_I  = \beta \delta_{Ii} x^i\,,
\end{split}
\end{equation}
where $f(r)$ is  the emblackening factor and $r_h$ denotes the black hole horizon. From here we set $L=1$.

From the solution \eqref{axion}, the thermodynamic quantities read as follows. $\mu$ is interpreted as the chemical potential of the boundary field theory, $\mu = \lim_{r\rightarrow \infty} A_t$. 
The temperature is
\begin{equation}
T=  \frac{f'(r_h)}{4\pi}  = \frac{1}{4\pi} \left( 3 r_h - \frac{\mu^2 +2  \beta^2}{4r_h}    \right) \,,
\end{equation}
so $r_h$ is expressed in terms of $T, \mu$ and $\beta$:
\begin{equation} \label{rh1}
r_h = \frac{1}{6}\left(4\pi T+\sqrt{6\beta^2+16\pi^2 T^2+3\mu^2}\right)\,.
\end{equation}
The entropy density is 

\begin{equation} \label{s1}
s =  4\pi r_h^2 = \frac{\pi}{9} \left(4\pi T+\sqrt{6\beta^2+16\pi^2 T^2+3\mu^2} \right)^2\,,
\end{equation}
by the Bekenstein-Hawking formula and the expectation value of the charge density reads
\begin{equation} \label{rho1}
\rho =  \mu r_h = \frac{\mu}{6} \left(4\pi T+\sqrt{6\beta^2+16\pi^2 T^2+3\mu^2} \right)\,.
\end{equation}
The butterfly velocity \eqref{BV1}  is
\begin{equation} \label{bv1}
v_B^2=  \frac{\pi T}{r_h}=\frac{6\pi T}{4\pi T+\sqrt{6\beta^2+16\pi^2 T^2+3\mu^2}}\,,
\end{equation}
 from the metric \eqref{axion} and \eqref{rh1}. 
 
The thermodynamic susceptibilities \eqref{sus1} are computed by using \eqref{s1} and \eqref{rho1} as follows. 
The compressibility and the thermoelectric susceptibility are
\begin{equation} \label{ts1}
\begin{split}
\chi &= \left( \frac{\partial \rho}{\partial \mu} \right)_{T} = \frac{1}{6} \left(4\pi T+\frac{6\beta^2+16\pi^2 T^2+6\mu^2}{\sqrt{6\beta^2+16\pi^2 T^2+3\mu^2}} \right)\,, \\
\zeta &= \left( \frac{\partial \rho}{\partial T} \right)_{\mu}=\left( \frac{\partial s}{\partial \mu} \right)_{T} = \frac{2\pi\mu}{3} \left(1+\frac{4\pi T}{\sqrt{6\beta^2+16\pi^2 T^2+3\mu^2}} \right)\,.
\end{split}
\end{equation}
The specific heat at fixed chemical potential or fixed charge density are
\begin{equation} \label{ts2}
\begin{split}
c_{\mu} &= T \left( \frac{\partial s}{\partial T} \right) = \frac{8\pi^2 T \left(4\pi T+\sqrt{6\beta^2+16\pi^2 T^2+3\mu^2} \right)^2}{9\sqrt{6\beta^2+16\pi^2 T^2+3\mu^2}}\,, \\
c_{\rho} &= c_\mu - \frac{\zeta^2 T}{\chi} = \frac{8\pi^2 T \left(4\pi T+\sqrt{6\beta^2+16\pi^2 T^2+3\mu^2} \right)^3}{9(6\beta^2+6\mu^2+4\pi T(4\pi T+\sqrt{6\beta^2+16\pi^2 T^2+3\mu^2}))}\,.
\end{split}
\end{equation}

To compute the conductivity we use the general formula \eqref{gencond}. In our model \eqref{eq:action}
\begin{align}
\phi=0, \quad \Phi_I=1, \quad Z=1, \quad V=-6, \quad \beta_I=\beta\,,
\end{align}
so the electrical, thermal, thermoelectric conductivities are
\begin{equation} \label{tc1}
\begin{split}
\sigma &= 1+\frac{\mu^2}{\beta^2}\,, \\
\overline \kappa &= \frac{4\pi s T}{\beta^2} = \frac{4\pi^2T}{9\beta^2} \left(4\pi T+\sqrt{6\beta^2+16\pi^2 T^2+3\mu^2} \right)^2\,, \\
\alpha &= \frac{4\pi \rho}{\beta^2} = \frac{2\pi\mu}{3\beta^2} \left(4\pi T+\sqrt{6\beta^2+16\pi^2 T^2+3\mu^2} \right)\,,
\end{split}
\end{equation}
and
\begin{equation} \label{tc2}
\kappa = \overline \kappa - \frac{\alpha^2 T}{\sigma} = \frac{4\pi^2T}{9(\beta^2+\mu^2)} \left(4\pi T+\sqrt{6\beta^2+16\pi^2 T^2+3\mu^2} \right)^2\,.
\end{equation}

\begin{figure}[]
	\begin{center}
		     { \subfigure[ Diffusion constants ($2\pi T D_{\pm}$) and the butterfly velocity ($v_B^2$) ]
			{\includegraphics[width=5cm]{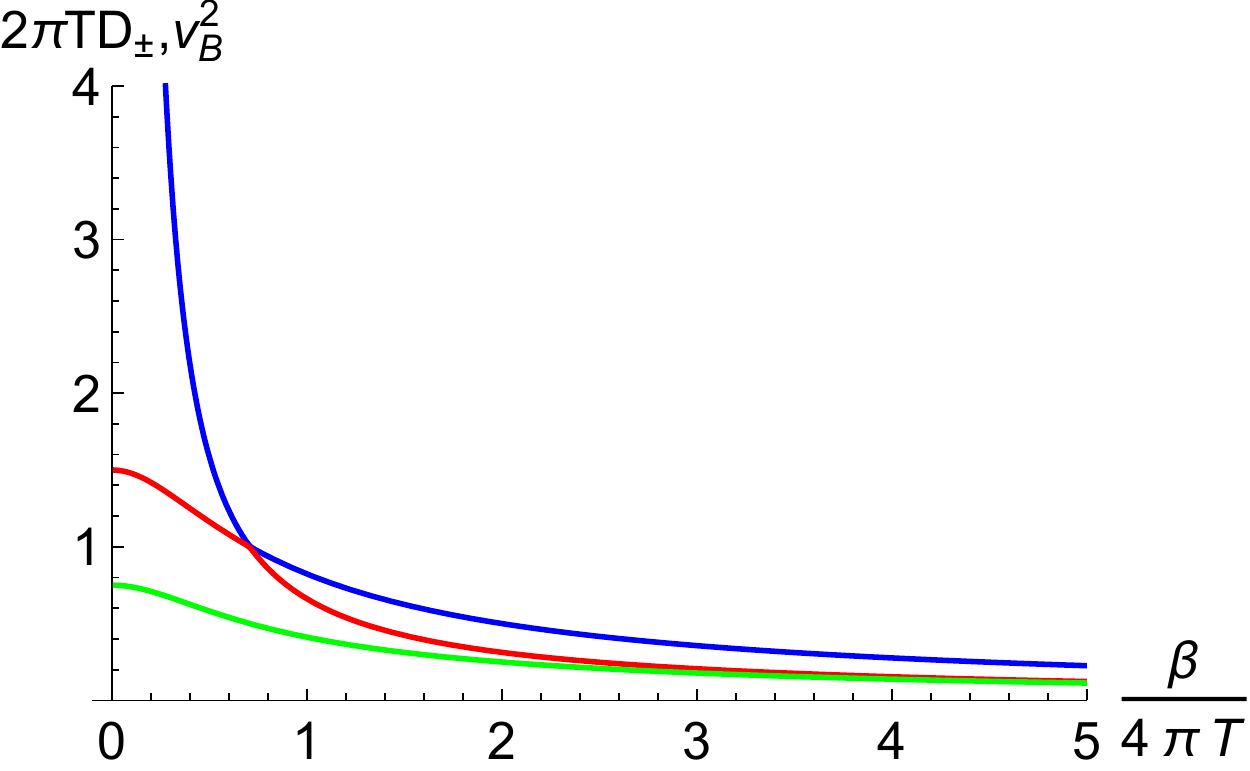}
			\includegraphics[width=5cm]{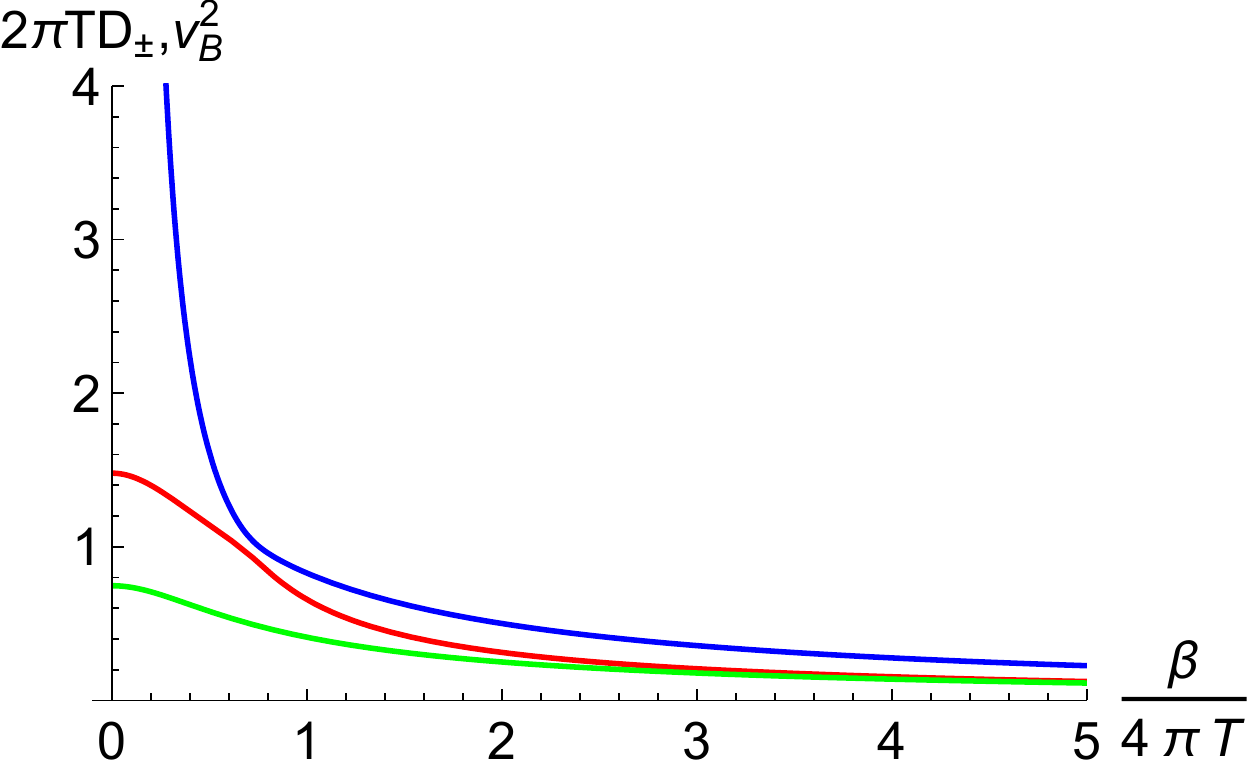}
			\includegraphics[width=5cm]{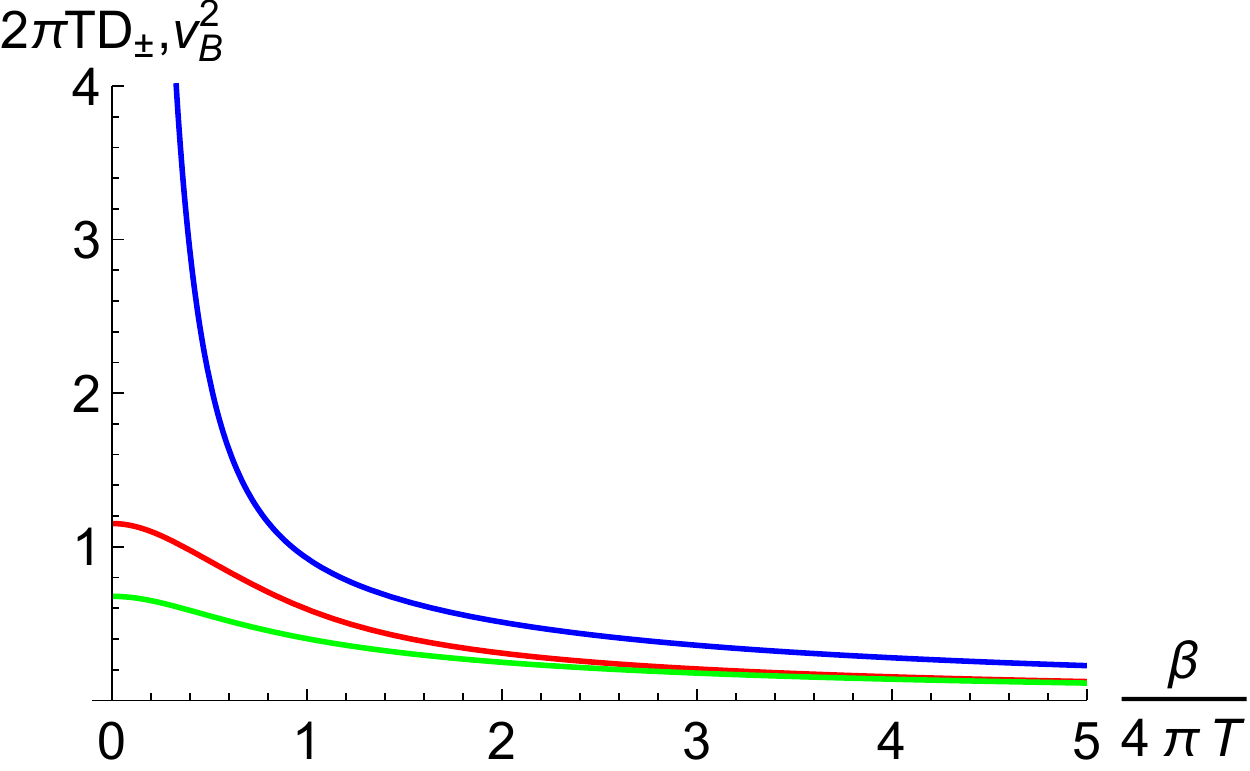}
			}
	              \subfigure[ Diffusion constants/(butterfly velocity)$^2$ ($C_\pm = 2\pi T D_{\pm}/v_B^2$)   ]
			{\includegraphics[width=5cm]{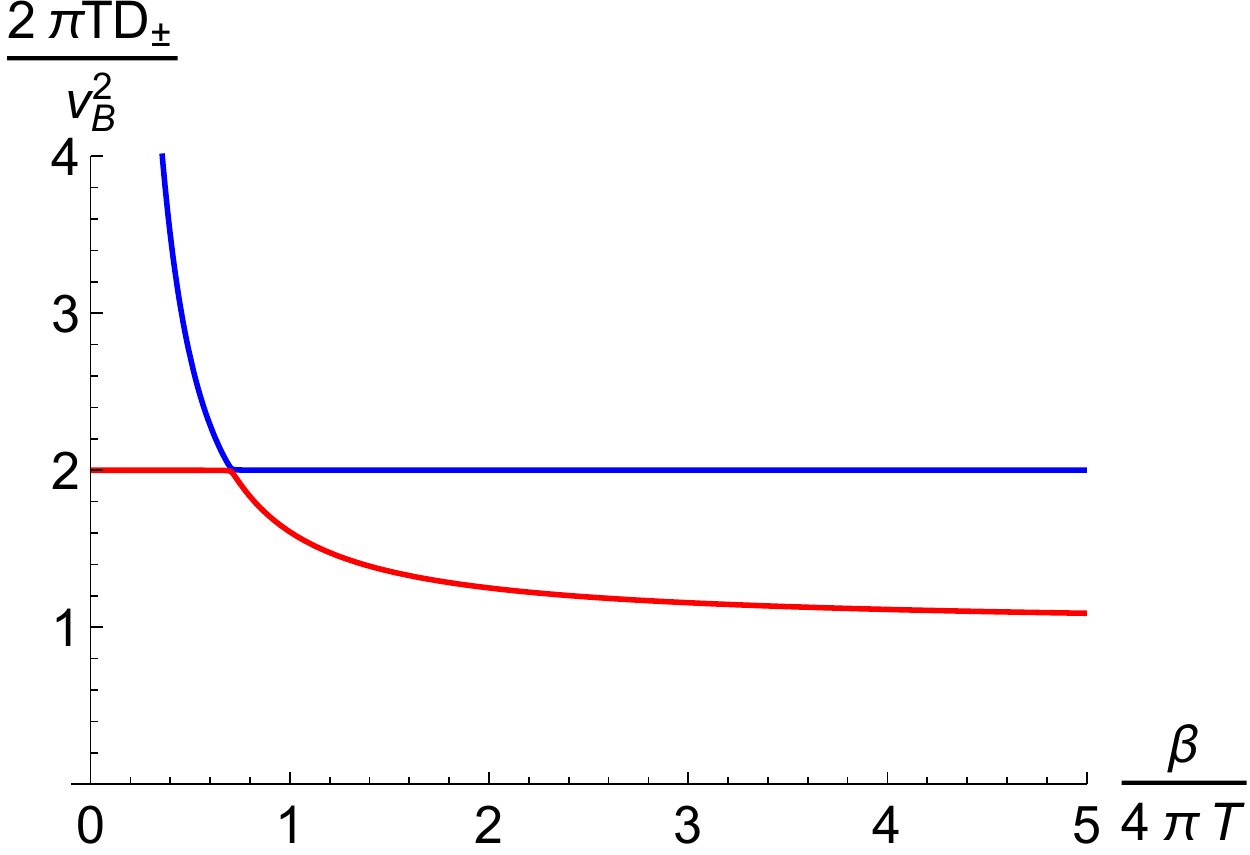}
			\includegraphics[width=5cm]{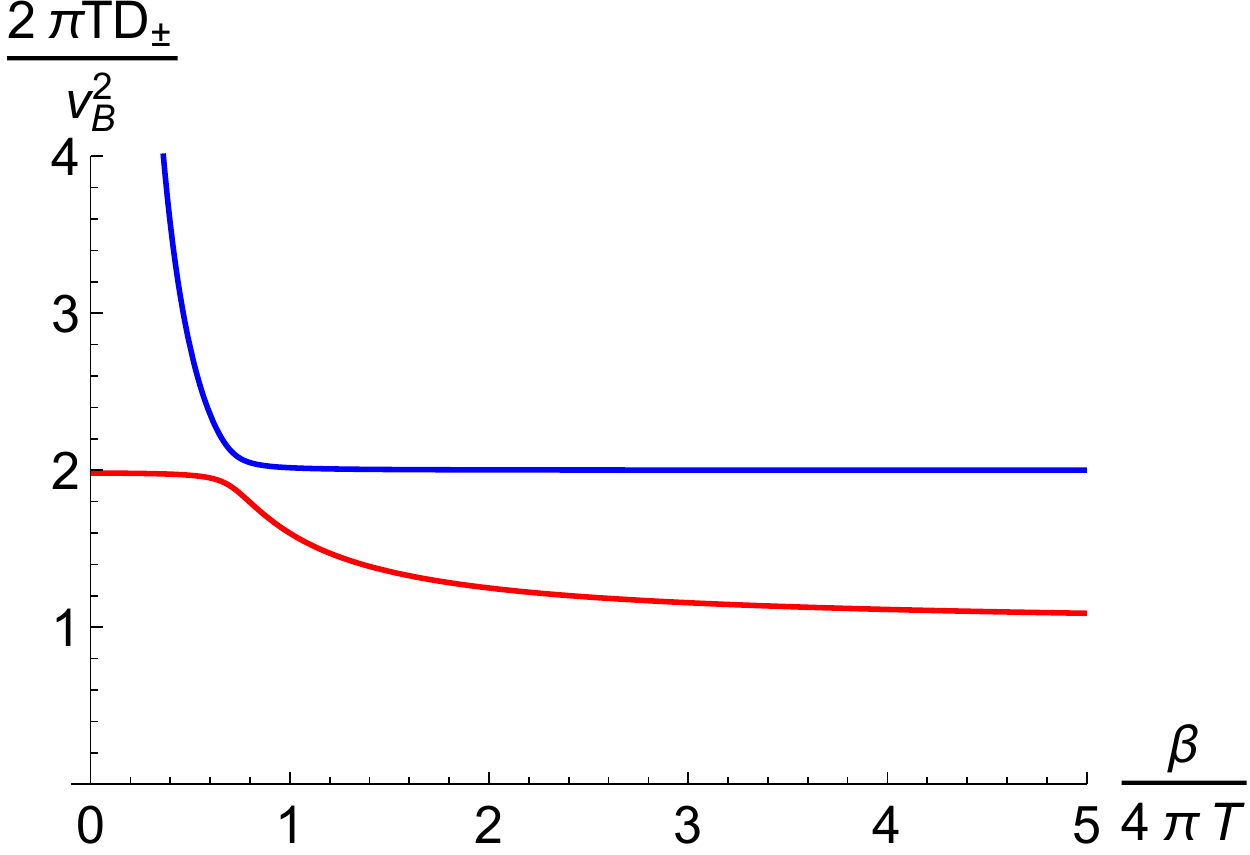}
			\includegraphics[width=5cm]{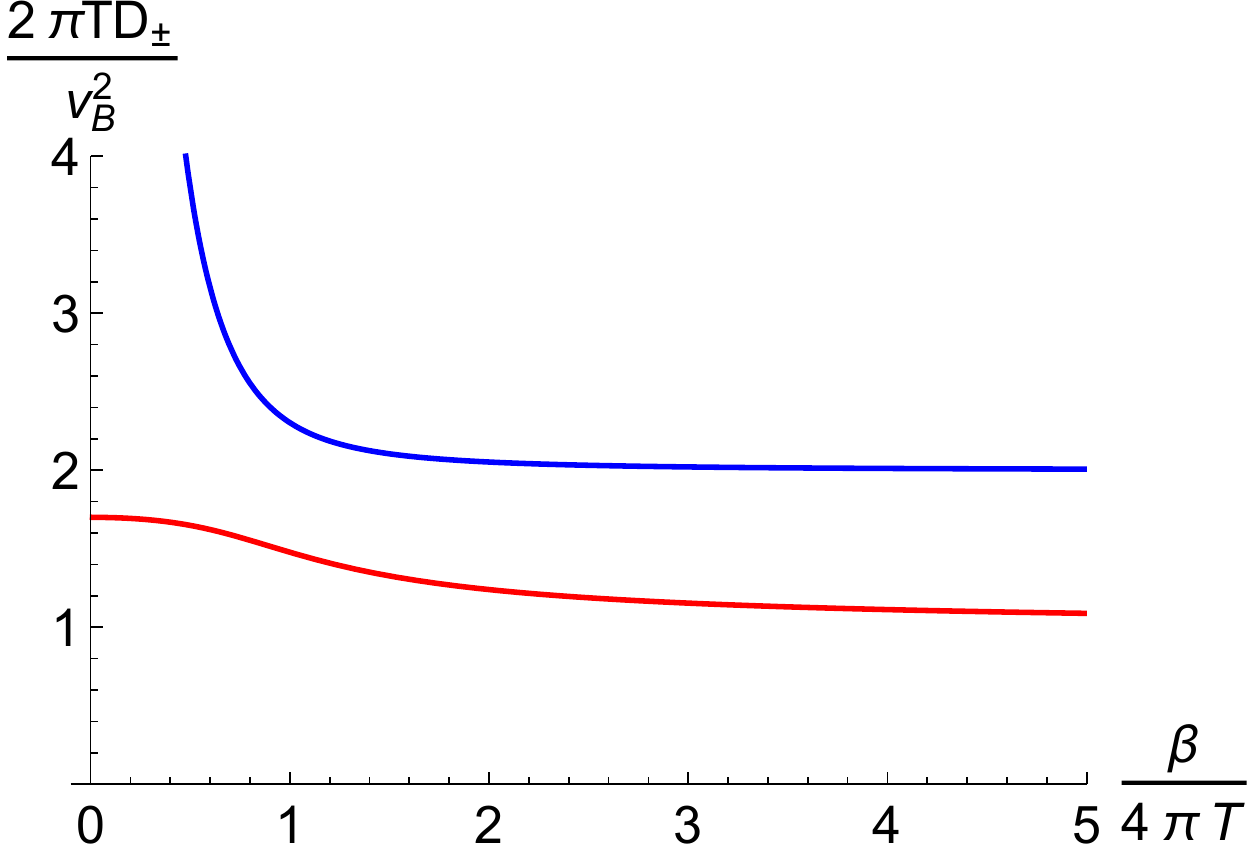}}}
	\end{center}
	\caption{Diffusion constants of the linear axion model at finite density. $\mu/T=0.1, 1, 5$ from left to right. The blue curve is for $D_+$ and the red curve is for $D_-$. The green curve displays $v_B^2$.  $2\pi D_{\pm}/v_B^2$ saturates the universal value in the incoherent regime:  $\beta/T \gg 1$ and $\beta/\mu \gg 1$.  } 	\label{fig1}
\end{figure}

Now we are ready to compute diffusion constants $D_\pm$, \eqref{DDD0} and \eqref{DDD1}, which we rewrite here for convenience, 
\begin{equation}
D_\pm = \frac{c_2 \pm \sqrt{c_2^2 - 4c_1}}{2} \,,
\end{equation}
where 
\begin{equation}
 c_1 = \frac{\sigma}{\chi} \frac{\kappa}{c_\rho}  \,,  \qquad  c_2= \frac{\sigma}{\chi} + \frac{\kappa}{c_\rho} + \mathcal{M} \,, \qquad 
\mathcal{M} =  \frac{T(\zeta \sigma - \chi \alpha)^2}{c_\rho \chi^2 \sigma}  \label{c1c22}  \,. 
\end{equation}
The analytic formulas of $D_\pm$  can be obtained by plugging \eqref{ts1}, \eqref{ts2}, \eqref{tc1}, and \eqref{tc2} into \eqref{c1c22}. Because the final expressions are complicated and not very illuminating  we show their plots in Fig. \ref{fig1}(a). 
The blue curve displays $2\pi T D_+$ and the red curve displays $2\pi D_-$. The green curve means $v_B^2$ \eqref{bv1}. As $\beta/T$ increases both $2\pi TD_{\pm}$ and $v_B^2$ go to zero, but all of them behave as  $1/(\beta/T)$. Thus, $2\pi TD_{\pm}/v_B^2$ saturate the finite lower bound as shown in Fig. \ref{fig1}(b).

These bounds in the incoherent regime ($\beta/T \gg 1$ and $\beta/\mu \gg 1$) can be read also from the analytic expression of $D_\pm$ and $v_B^2$ at large $\beta$:
\begin{align}
D_+ &= \frac{\sqrt{6}}{\beta }-\frac{4 \pi  T}{\beta ^2}+\frac{3 \sqrt{6} \mu ^2+16 \pi ^2 \sqrt{6} T^2}{12 \beta ^3} \cdots \,, \\
D_- &=  \frac{\sqrt{\frac{3}{2}}}{\beta }+\frac{16 \sqrt{6} \pi ^2 T^2-3 \sqrt{6} \mu ^2}{24 \beta ^3} \cdots \,, \\
v_B^2 &= \sqrt{6}\pi \frac{ T}{\beta }-\frac{4 \pi ^2 T^2}{\beta ^2}+\frac{16 \sqrt{6} \pi ^3 T^3-3 \sqrt{6} \pi  \mu ^2 T}{12 \beta ^3} \,,
\end{align}
which yield
\begin{align}
C_+ &= \frac{2\pi T D_+}{v_B^2}  =  2+  \frac{\mu ^2}{\beta ^2}+\frac{2 \pi  \sqrt{\frac{2}{3}} \mu ^2 T}{\beta ^3} + \cdots \,,  \\
C_-  &= \frac{2\pi T D_-}{v_B^2}   =1+\frac{2 \pi  \sqrt{\frac{2}{3}} T}{\beta } +\frac{8 \pi ^2 T^2}{3 \beta ^2} + \frac{16 \sqrt{6} \pi ^3  T^3-9 \sqrt{6} \pi  \mu ^2 T}{18 \beta ^3} + \cdots
\,.
\end{align}
We find that $C_\pm$ saturate the bounds  $C_+ = 2$ and $C_- = 1$  in the incoherent regime. 
Indeed, in this regime, $D_+$ and $D_-$ can be identified with $D_c$ and $D_e$ respectively because the mixing term $\mathcal{M}$ in \eqref{c1c22} vanishes  as explained in the following paragraph.  
Thus the bounds  $C_+ = 2$ and $C_- = 1$ at zero density \cite{Blake:2016sud} still hold at finite density in the incoherent regime.

%
%\begin{equation}
%\frac{4 \sqrt{\frac{2}{3}} \pi ^2 \mu ^2 T^2}{\beta^5}
%\end{equation}
%

%%%%%%%%%%%%%%%%%%%%%%%%%%

\begin{figure}[]
	\begin{center}
		      \subfigure[ Diffusion constants: $2\pi T D_c$ and $2\pi T D_e$ ]
			{
			\includegraphics[width=5cm]{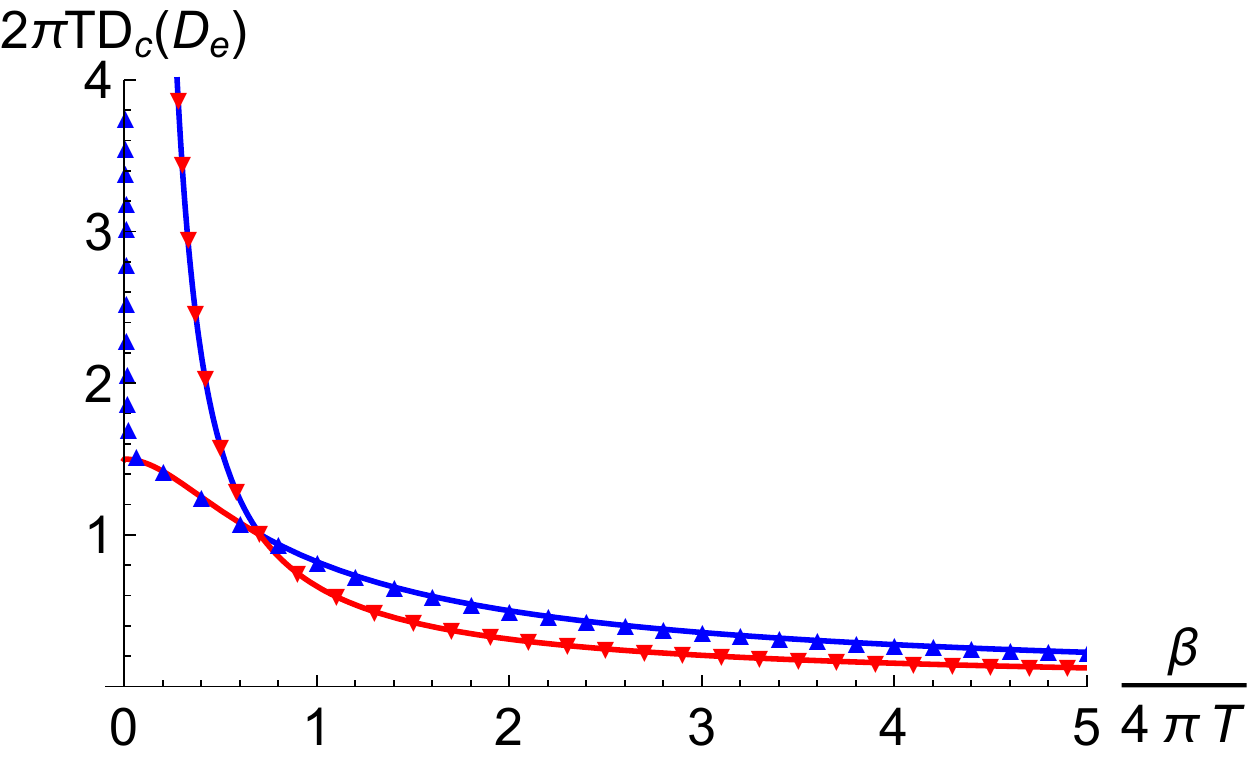}
			\includegraphics[width=5cm]{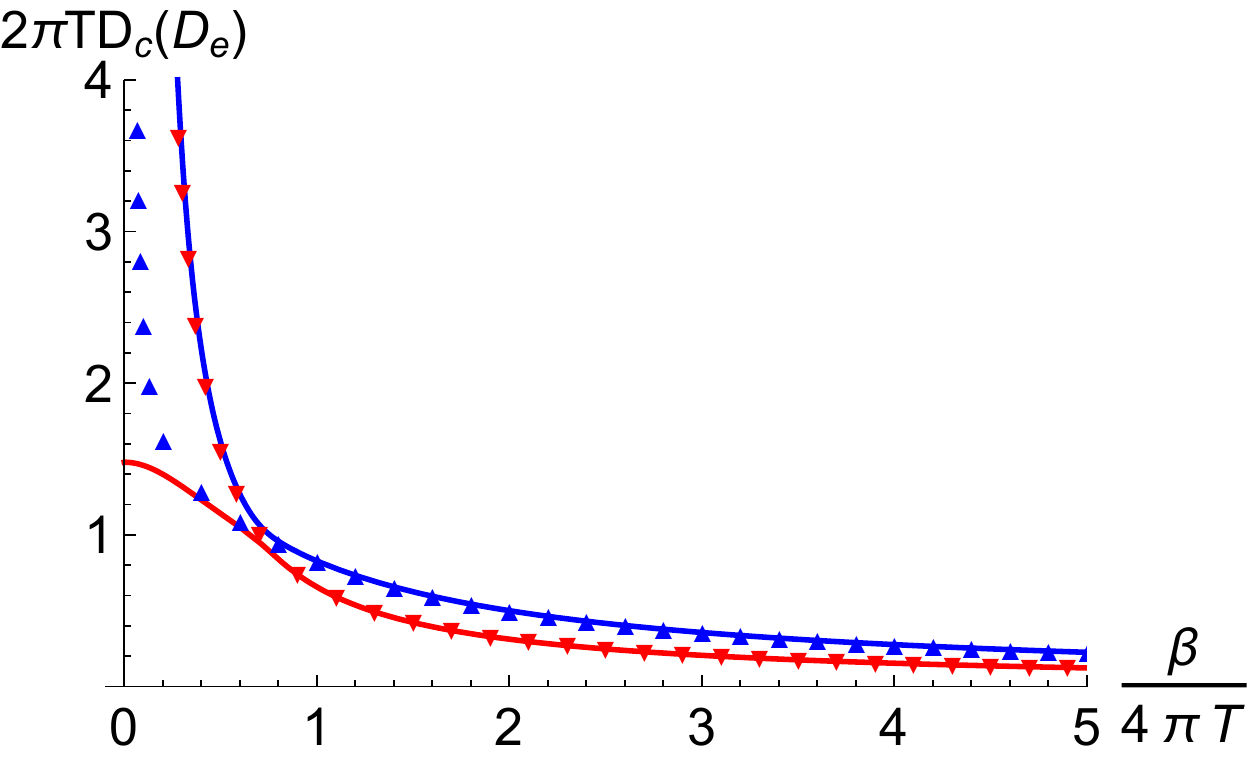}
			\includegraphics[width=5cm]{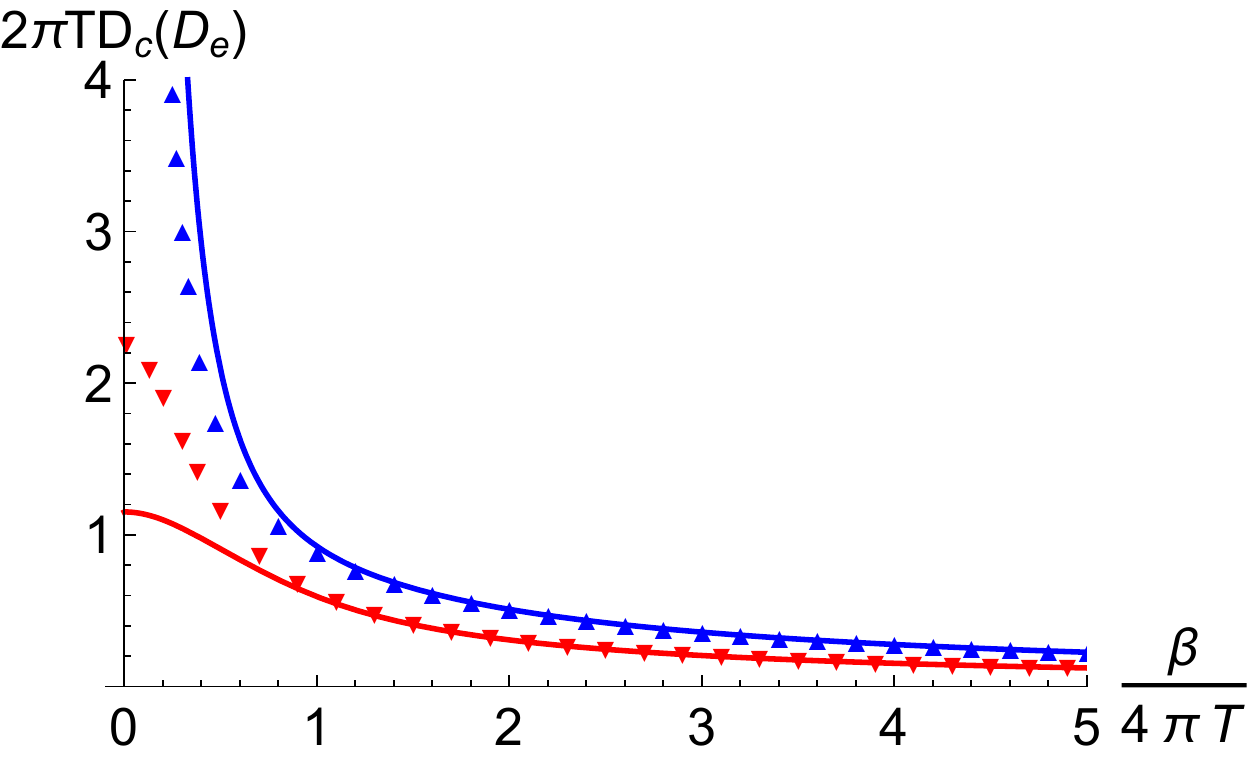}
			}
	              \subfigure[ Diffusion constants/(butterfly velocity)$^2$: $2\pi T D_c/v_B^2$ and $2 \pi T D_e/v_B^2$ ]
			{
			\includegraphics[width=5cm]{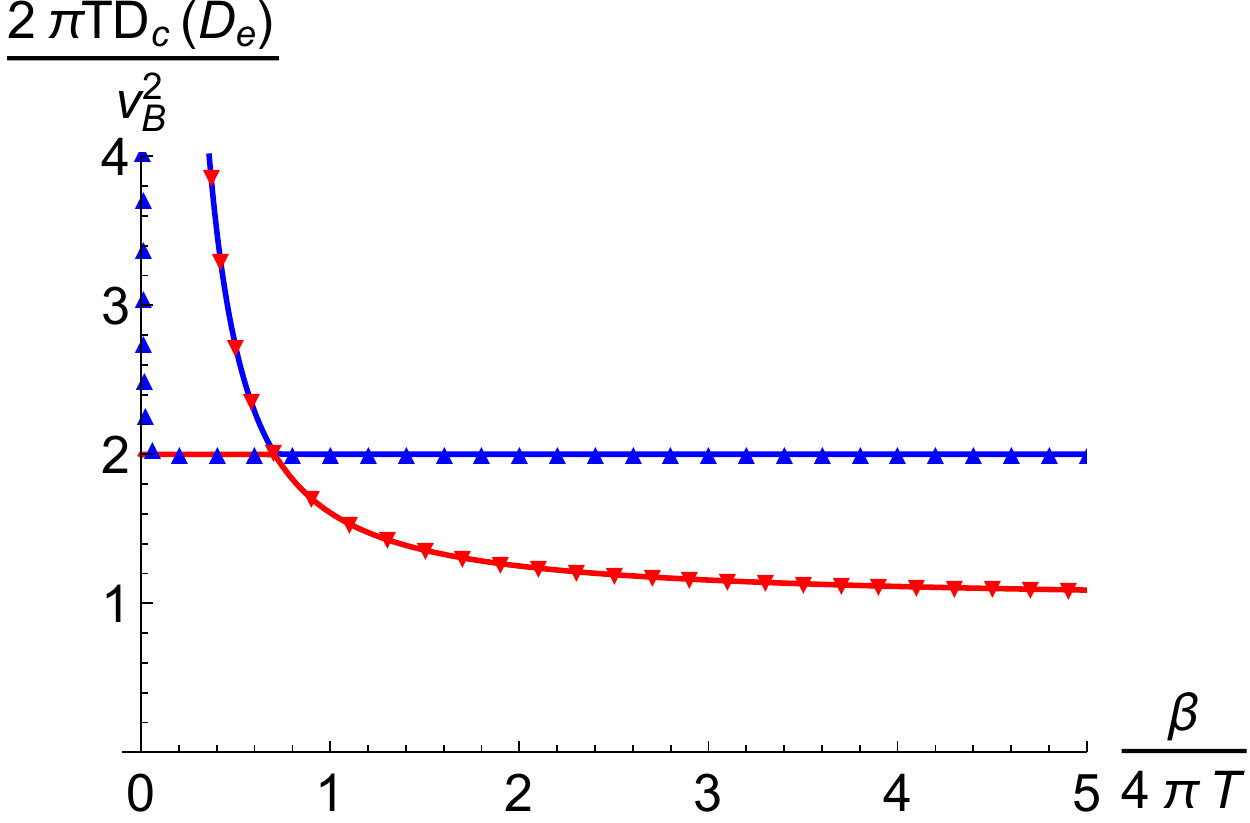}
			\includegraphics[width=5cm]{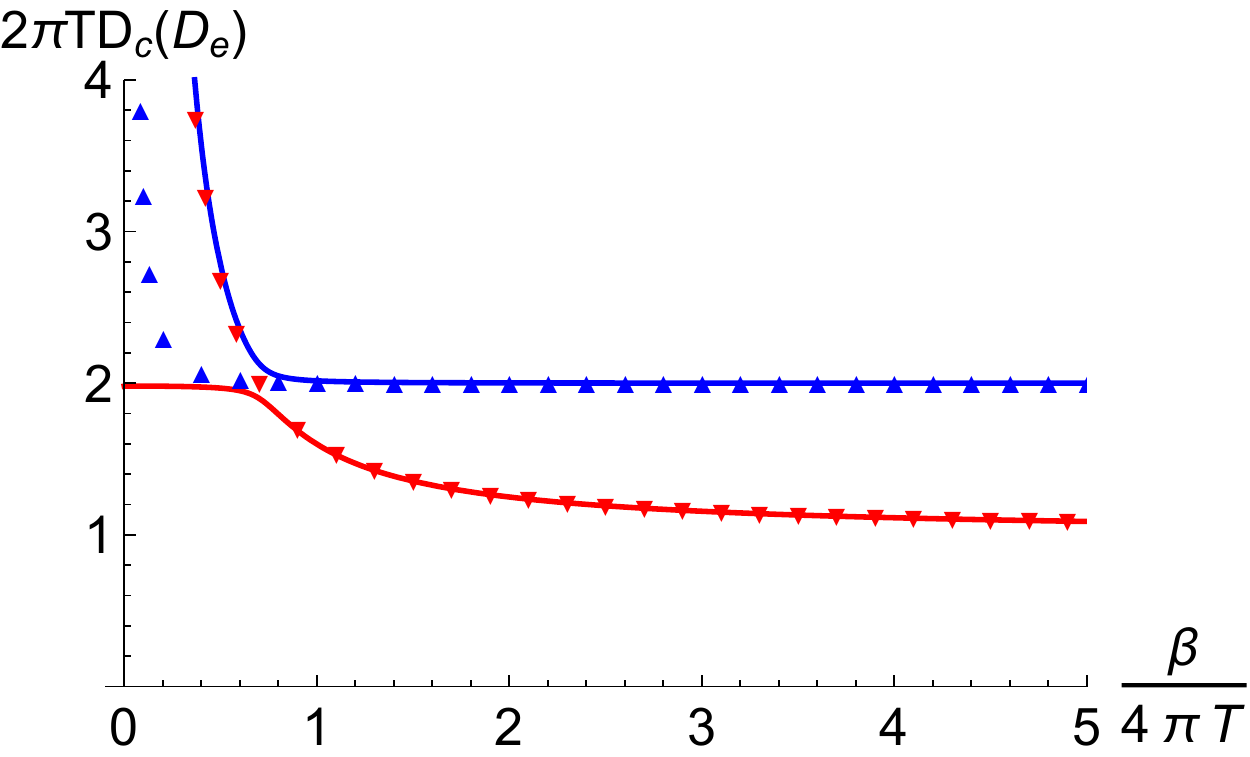}
			\includegraphics[width=5cm]{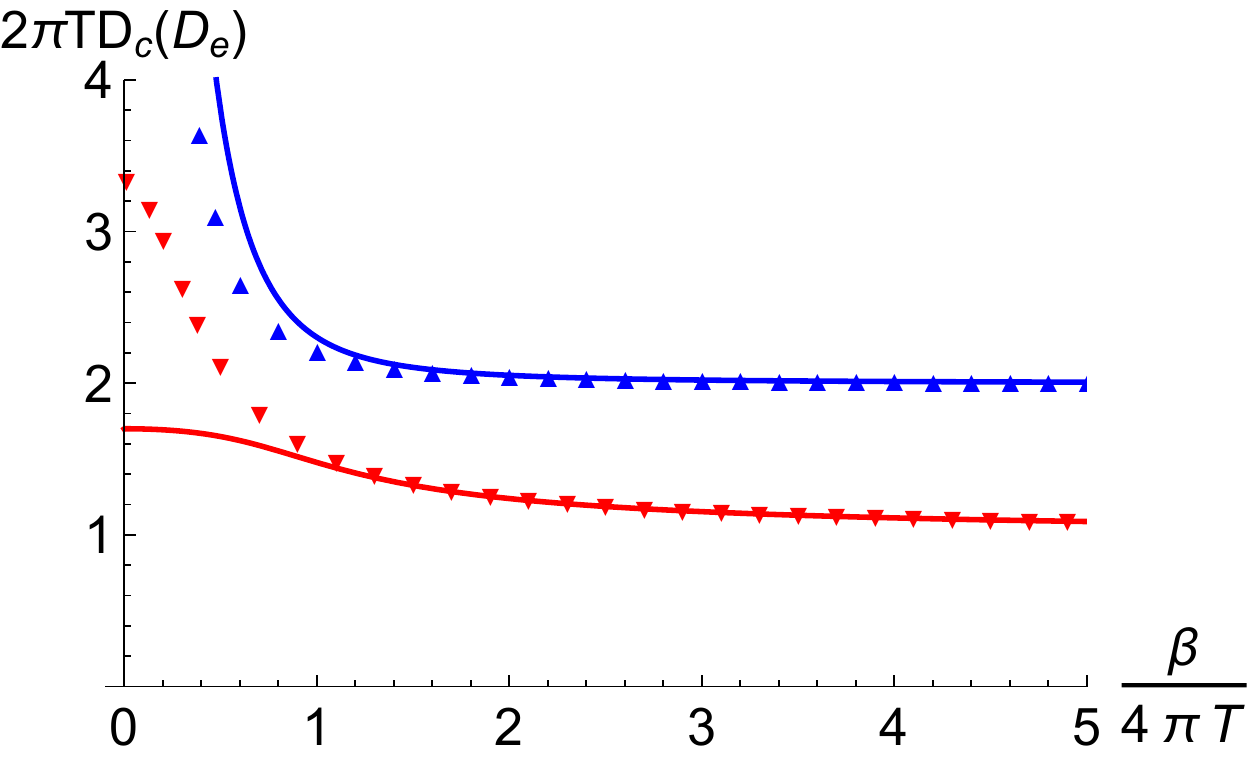}}
	\end{center}
	\caption{Diffusion constants of the linear axion model with/without a mixing term. $\mu/T=0.1, 1, 5$ from left to right. 
	The blue curve is for $D_c$ and the red curve is for $D_e$. The triangles display the results without the mixing term $\mathcal{M}$ in  \eqref{c1c22}. For comparison, we also display  the results  with the mixing term, the solid curves in Fig. \ref{fig1}. }	\label{fig2}
\end{figure} 
%%%%%%%%%%%%%%%%%%%%%%%%%%

To see the effect of the mixing term  $\mathcal{M}$ in \eqref{c1c22}, we may drop  it in \eqref{c1c22} and simply identify $D_c \equiv \sigma/\chi$ and $D_c \equiv \kappa/c_\rho$. These are shown as the triangles in Fig. \ref{fig2}, where the solid curves in Fig. \ref{fig1} are displayed together for comparison. For $\beta/4\pi T \lesssim 1 $ the mixing term is important and for $\beta/4\pi T \gtrsim 1$ it is negligible. For small $\mu/T=0.1$, this mixing effect is `maximal' in the sense $D_-$ is identified with $D_c$ and $D_+$ is identified with $D_e$, which is opposite to the case in the incoherent regime.  However, even for small $\beta/T$, the effect of the mixing term decreases as $\mu/T$ increases. The effect of $\mu/T$ also can be checked by  large $\mu/T$  expansion of $T\mathcal{M}$:
\begin{equation}
T\mathcal{M} = \frac{4 \pi ^2 T^3}{\sqrt{3} \mu  \beta^2} + \cdots \,,
\end{equation}
which implies that the mixing term may be small if $\mu/T$ is large even for small $\beta/T$. 
In conclusion, in the incoherent regime, the mixing term, $\mathcal{M}$ in  \eqref{c1c22}, between the charge and energy diffusions are negligible so $D_+$ and $D_-$ may be identified with $D_c$ and $D_e$ respectively. (Because $\mu/T \leqslant 5$ in Fig. \ref{fig2}, $\beta/4\pi T \sim 5$ is the incoherent regime.)

For $C_\pm$ to be universal it is important to consider the incoherent regime ($\beta/T \gg 1$ and $\beta/\mu \gg 1$). To see it more clearly let us consider a different case $\mu/T \gg 1$ and $\mu/\beta \gg 1$, which yield the following expansions: 
\begin{equation}
D_+ = \frac{\sqrt{3} \mu }{\beta ^2}-\frac{2 \pi  T}{\beta ^2}, \cdots \,, \quad
D_- = \frac{\sqrt{3}}{\mu }-\frac{2 \pi  T}{\mu ^2}+\cdots \,, \quad
v_B^2 = \frac{2 \sqrt{3} \pi  T}{\mu }-\frac{8 \pi ^2 T^2}{\mu ^2}+\cdots \,,
\end{equation}
and
\begin{align}
C_+ &= \frac{2\pi T D_+}{v_B^2}  =  \frac{\mu ^2}{\beta ^2}+\frac{2 \pi  \mu  T}{\sqrt{3} \beta ^2}, + \cdots \,,  \\
C_-  &= \frac{2\pi T D_-}{v_B^2}   = 1+ \frac{2 \pi  T}{\sqrt{3} \mu } + \cdots
\,.
\end{align}
Thus $C_+$ is not universal while $C_-$ is.

Finally, let us compare our results with \cite{Blake:2016jnn} where {\it low temperature limit} was considered in a class of holographic models that flow to $AdS_2 \times R^d$ fixed points in the infrared. Our model belongs to that class and serves as a concrete example. Low temperature expansion of our solution is as follows. 
\begin{align}
D_+ &= \frac{\sqrt{3} \sqrt{2 \beta ^2+\mu ^2}}{\beta ^2}-\frac{2 \pi   \left(2 \beta ^2+\mu ^2\right)}{\beta ^2 \left(\beta ^2+\mu ^2\right)} T + \cdots \,, \\
D_- &=\frac{\sqrt{3}}{\sqrt{2 \beta ^2+\mu ^2}}-\frac{2 \pi  \mu ^2 }{2 \beta ^4+3 \beta ^2 \mu ^2+\mu ^4} T +\cdots \,, \\
v_B^2 &= \frac{2 \sqrt{3} \pi  }{\sqrt{\mu ^2+2 \beta^2}} T + \cdots \,,
\end{align}
where $\sqrt{2 \beta ^2+\mu ^2}/\sqrt{3}$ corresponds to  $c_h^0$ in \cite{Blake:2016jnn}. The mixing term between the charge and energy diffusion is expanded as 
\begin{equation}
\mathcal{M} = \frac{\mu^2 }{ \beta^2}\frac{4 \pi ^2  \sqrt{2 \beta^2+\mu^2}}{\sqrt{3} \left(\beta^2+\mu ^2\right)^2} T^2 + \cdots \,,
\end{equation}
which can be ignored at low temperature as argued in \cite{Blake:2016jnn}. Thus we may identify $D_+ = D_c$ and $D_- = D_e$ and
\begin{align}
C_+ = \frac{2\pi T D_c}{v_B^2} &= \left(2+ \frac{\mu ^2}{\beta^2}\right)+ \frac{\mu^2}{\beta^2}\frac{2 \pi   \sqrt{2 \beta ^2+\mu ^2}}{\sqrt{3} \left(\beta ^2+\mu ^2\right)}T + \cdots  \,,  \\
C_ - = \frac{2\pi T D_e}{v_B^2} &= 1+ \frac{2 \pi   \sqrt{2 \beta ^2+\mu ^2}}{\sqrt{3} \left(\beta ^2+\mu ^2\right)}T + \cdots  \,.
\end{align}
The thermal diffusion constant has a universal coefficient $C_-$ irrespective of $\beta$ and $\mu$, which agree to \cite{Blake:2016jnn}. The coefficient $C_+ = 2+\mu^2/\beta^2$ for charge diffusion constant is not universal and a function of $\mu$ and $\beta$, but in the incoherent regime ($\mu/\beta \ll 1$) it becomes universal. i.e. $C_+ = 2$.

\section{Axion-dilaton model}\label{sec4}

Next, let us consider a four dimensional Einstein-Maxwell-Axion-Dilaton theory. It is based on the Einstein-Maxwell-Dilaton model so called the Gubser-Rocha model~\cite{Gubser:2009qt}. To include momentum relaxation effect, a graviton mass term breaking translational invariance was added to the action and the linear-$T$-resistivity was observed in this model~\cite{Davison:2013txa}. Because the universal bound of the diffusion constant may be related to the 
linear-$T$-resistivity it will be interesting to investigate the diffusion constants in this model. 
Here, we  modify the original Gubser-Rocha model by adding the scalar fields $\psi_I$ instead of a graviton mass term to induce the momentum relaxation\footnote{The conductivities of this model was also studied in \cite{Zhou:2015qui}, focusing on the slow momentum relaxation in low frequency approximation.}. It makes possible a direct comparison with the linear axion model in the previous section.

The action is
\begin{equation}
S=\int \mathrm{d}^4x\sqrt{-g} \left[R-\frac{1}{4} e^\phi F^2 -\frac{3}{2}(\partial{\phi})^2+\frac{6}{L^2}\cosh \phi -\frac{1}{2}\sum_{I=1}^{2}(\partial \psi_{I})^2  \right] \,,
\end{equation}
which belongs to \eqref{genact} and is reduced to \eqref{eq:action} if $\phi = 0$.
An analytic classical solution is 
\begin{equation} \label{GReq}
\begin{split}
\mathrm{d} s^2 &=\frac{r^2 g(r)}{L^2} \left(-h(r) \mathrm{d} t^2+\mathrm{d} x^2 +\mathrm{d} y^2 \right)+\frac{L^2}{r^2 g(r)h(r)}\mathrm{d} r^2 \,,\\
& \quad h(r)=1-\frac{L^4 \beta^2}{2(Q+r)^2}-\frac{(Q+r_h)^3}{(Q+r)^3}\left(1-\frac{L^4 \beta^2}{2(Q+r_h)^2}\right) \,,  \quad g(r)=\left(1+\frac{Q}{r}\right)^\frac{3}{2} \,, \\
%\mathrm{d} s^2 &=\frac{r^2 g(r)}{L^2} \left(-h(r) \mathrm{d} t^2+\mathrm{d} x^2 +\mathrm{d} y^2 \right)+\frac{L^2}{r^2 g(r)h(r)}\mathrm{d} r^2 \,,\\
%& \quad h(r)=1-\frac{L^4 \beta^2}{2(Q+r)^2}-\frac{(Q+r_h)^3}{(Q+r)^3}\left(1-\frac{L^4 \beta^2}{2(Q+r_h)^2}\right) \,,  \quad g(r)=\left(1+\frac{Q}{r}\right)^\frac{3}{2} \,,\\
A&=\sqrt{\frac{3Q(Q+r_h)}{L^2}\left(1-\frac{L^4 \beta^2}{2(Q+r_h)^2}\right)}\left(1-\frac{Q+r_h}{Q+r}\right) \mathrm{d} t \,,\\
\phi&=\frac{1}{3} \log(g(r))\,,  \quad \psi_I =  \beta \delta_{Ii} x^i\,.
\end{split}
\end{equation}
where $r_h>0$ and $r_h>-Q$ to have a regular solution. Note that $Q$ can be negative. From here we set $L=1$.

\subsection{Thermodynamics and transport coefficients}

By the same method as in the linear-axion model, the temperature, the entropy density, chemical potential and charge density are
\begin{align} 
T &=  \frac{\left[r_h^2 g(r_h) h(r_h)\right]'}{4\pi}=  \frac{\sqrt{r_h}(6(Q+r_h)^2-\beta^2)}{8\pi (Q+r_h)^{3/2}} \,, \label{T2} \\
s  &= \frac{ r_h^2  g(r_h)}{4 G} =  4\pi\sqrt{r_h}(Q+r_h)^{3/2} \,, \label{s2} \\
\mu &=\sqrt{3Q(Q+r_h)\left(1-\frac{\beta^2}{2(Q+r_h)^2}\right)} \label{mu2} \,, \\
%&=\sqrt{\frac{6Q(Q+r_h)}{r_h\beta^2/T^2}} \sqrt{2 \sqrt{2} \pi  \sqrt{Q+r_h} 
%	\sqrt{8 \pi ^2 (Q+r_h)+3 r_h\beta^2/T^2}-8 \pi ^2
%	(Q+r_h)-r_h\beta^2/T^2}\,, \\
\rho & = (Q+r_h)\sqrt{3Q(Q+r_h)\left(1-\frac{\beta^2}{2(Q+r_h)^2}\right)}\,. \label{rho2}
\end{align}

 Thanks to a scaling symmetry of the solution, it is convenient to define the following variables:
\begin{equation}
\ttt = {t}\, {r_h}\,, \quad  \txx = {x}\, {r_h} \,,  \quad \tyy = {y}\, {r_h} \,,  \quad \trr = \frac{r}{r_h} \,, \quad \tbeta = \frac{\beta}{r_h} \,,
\quad \tQ = \frac{Q}{r_h} \,, 
\end{equation}
Consequently, we may define the field theory variables as:
\begin{equation}
\tTemp = \frac{T}{r_h} \,, \quad \tss = \frac{s}{r_h^2} \,, \quad \tmu = \frac{\mu}{r_h} \,, \quad \trho = \frac{\rho}{r_h^2} \,.
\end{equation}
In other words, we may set $r_h = 1$ in \eqref{GReq} and  replace all variables with the `tilde' variables.  In \eqref{GReq}  $\tQ$ and $\tbeta$ look natural independent variables but in our analysis, from the perspective of the dual field theory, $\mu/T$ and $\beta/T$ will be used as independent variables. By the relations \eqref{T2} and \eqref{mu2}, $\tQ$ and $\tbeta$ are expressed in terms of $\mu/T$  and $\beta/T$.

\begin{figure}
	\begin{center}
	    {  \subfigure[ $\mu/T=0.1$ ]
		{\includegraphics[width=6cm]{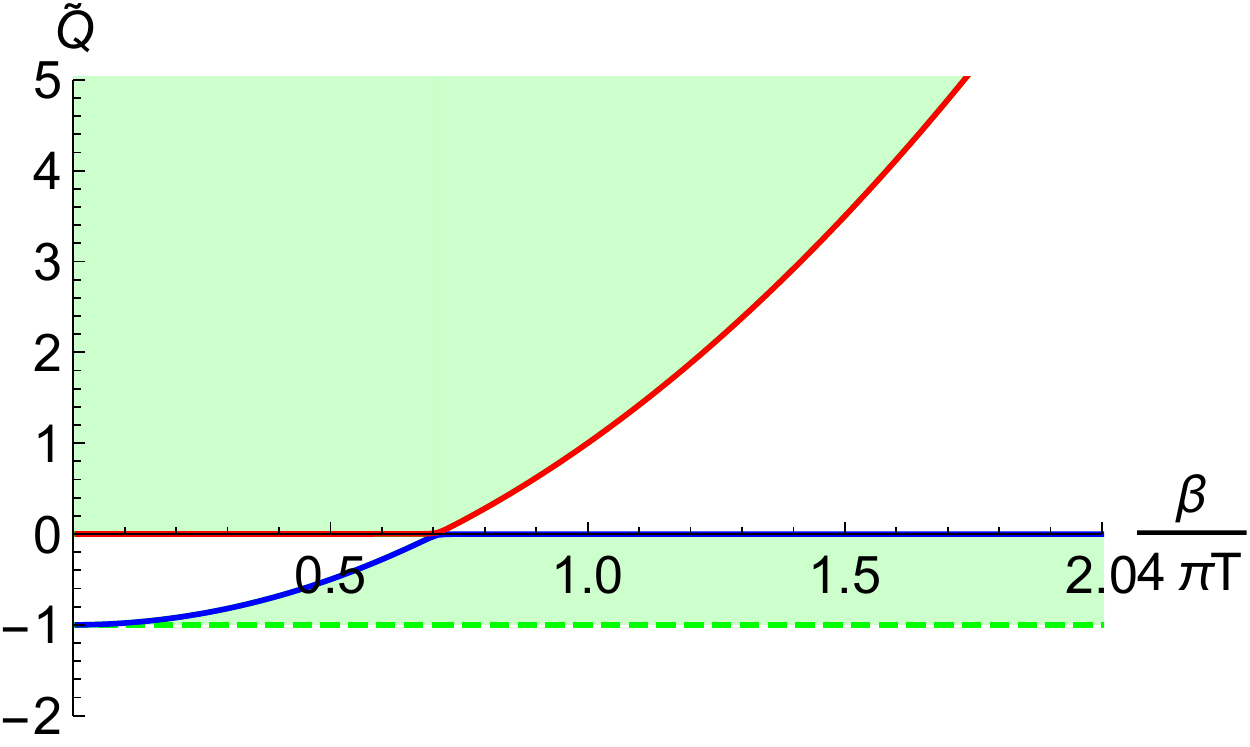}
		\includegraphics[width=6cm]{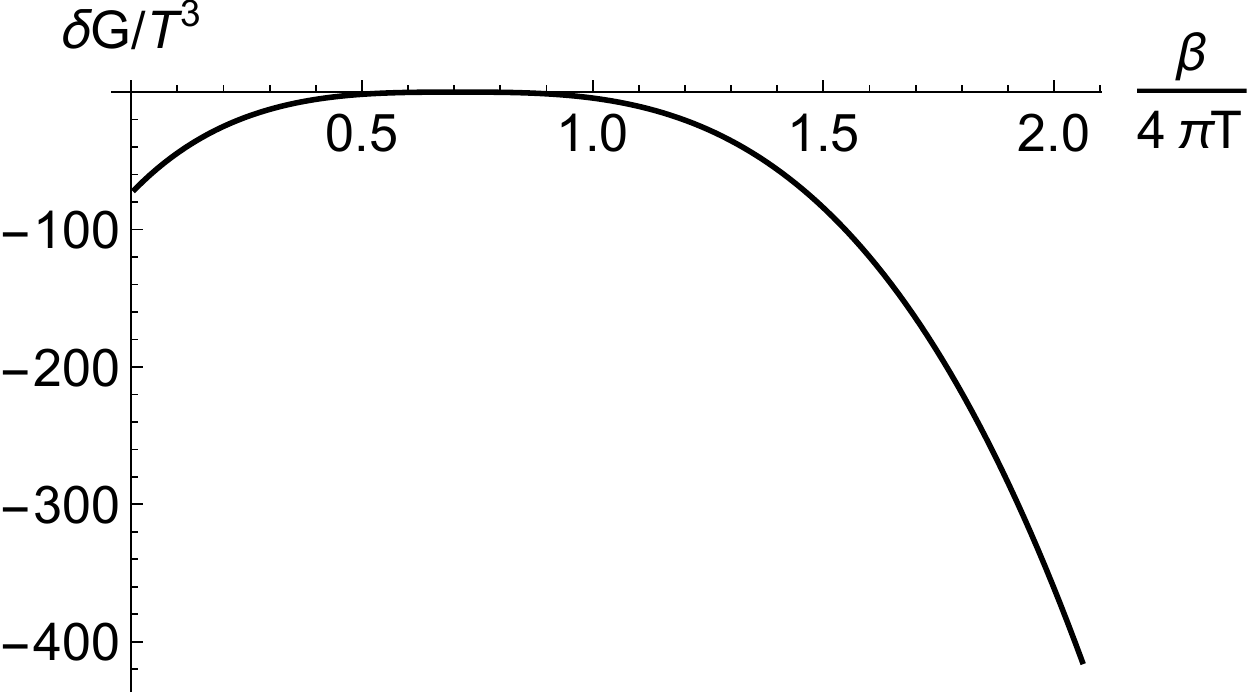}}
              \subfigure[ $\mu/T=5$ ]
		{\includegraphics[width=6cm]{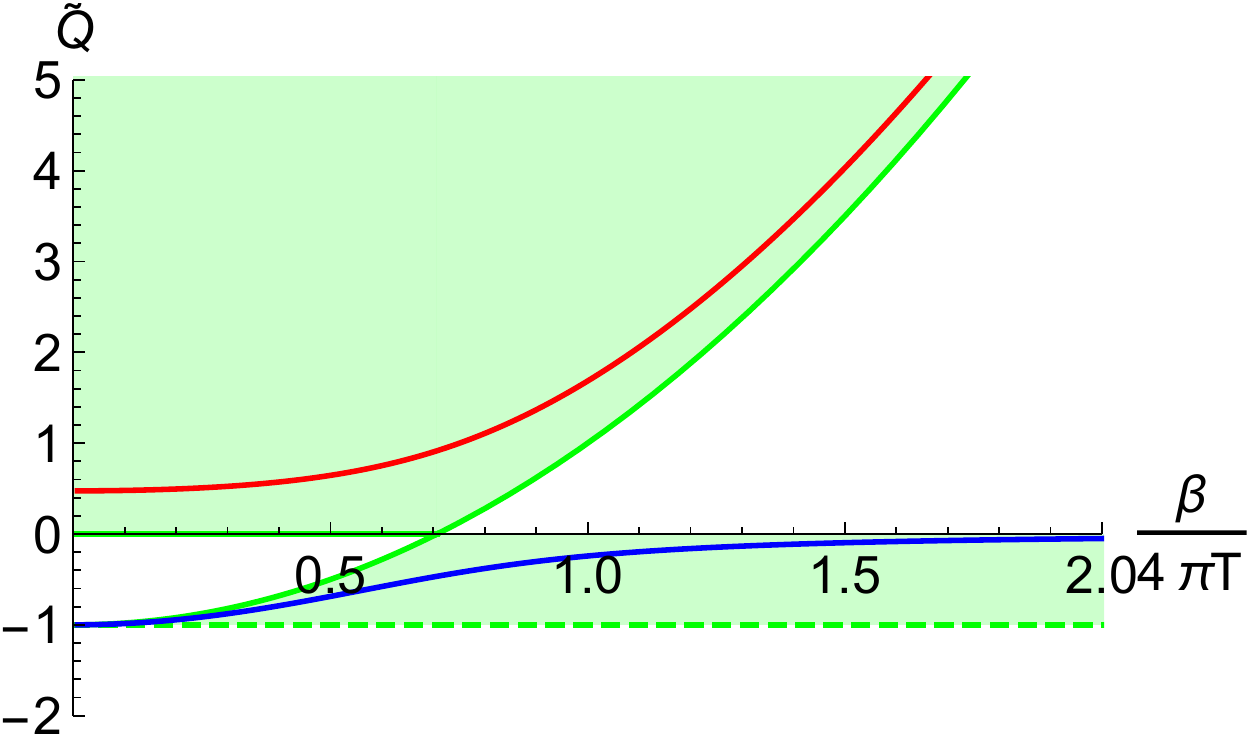}
		\includegraphics[width=6cm]{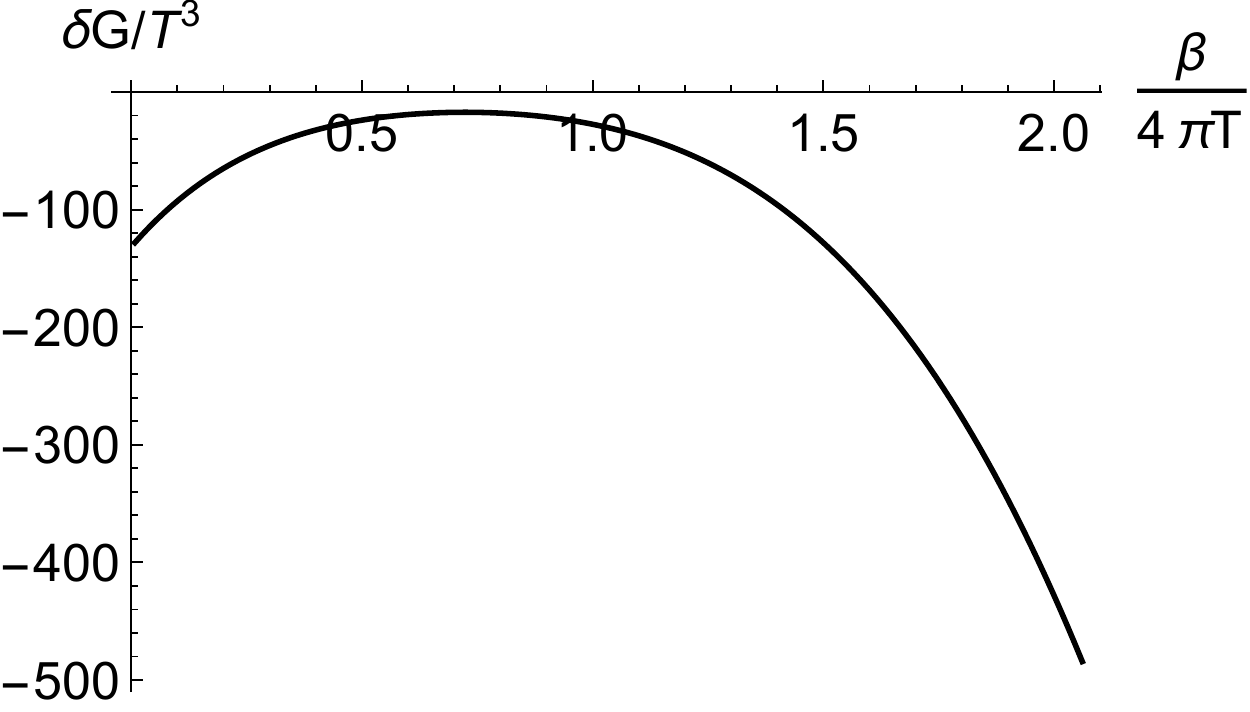}}}
	\end{center}
	\caption{The left panel: two values of $\tQ$ for given $\mu/T=0.1$(a) and $\mu/T=5$(b). The red curve is for positive $\tQ$ and the blue curve is for negative  $\tQ$.  The right panel: the difference of the grand potential ($\delta \grandpot = \grandpot(\tQ>0) - \grandpot(\tQ<0)$ is shown. The positive $\tQ$ solution is always thermodynamically preferred. The green region represent two branches in \eqref{branches}. }\label{tQregion}
\end{figure}

To find a possible range of $\tQ$, we impose physical condition $T \ge 0$, $\mu \ge 0$ and $\tQ > -1$. Without loss of generality,  we can take $\beta \ge 0$.  All these inequalities imply
	    \begin{equation} \label{branches}
	    \begin{cases}
	    \quad  \frac{\beta}{4\pi T}=0 &\Rightarrow \tQ \ge 0  \\
	    0<\frac{\beta}{4\pi T} \le \frac{1}{\sqrt{2}} &\Rightarrow
	        \begin{cases}
	        - 1 <\tQ \le  - 1+ 2 \left(\frac{\beta}{4\pi T}\right)^2 \\ 0 \le \tQ
	        \end{cases} \\	    
	    \quad  \frac{1}{\sqrt{2}} \le \frac{\beta}{4\pi T}  &\Rightarrow
	        \begin{cases}
	        -1<\tQ \le 0 \\ - 1 +  2 \left(\frac{\beta}{4\pi T}\right)^2  \le \tQ
	        \end{cases}
	    \end{cases} .
     	\end{equation}
Here we find that there may be two branches of the solutions: positive $\tQ$ and negative $\tQ$. 
They correspond to the green region in Fig. \ref{tQregion}.  The boundary of the green region is nothing but the condition for  
$\mu=0$. Indeed, by using \eqref{T2} and \eqref{mu2} we can obtain two solutions. For example, for $\mu/T=0.1$, and $5$, they are shown in Fig. \ref{tQregion}: the red curve for positive $\tQ$ and  the blue curve for negative $\tQ$. 
To determine which one corresponds to the ground state solution, we compare the grand potential density of two solutions. 
The grand potential density ($\grandpot$) is the on-shell action divided by the spatial volume and temperature \cite{Caldarelli:2016nni} 
\begin{align}
\grandpot& =\epsilon -Ts-\mu\rho\ \nonumber\\
&=-\frac{\beta^2Q}{2}+\frac{2\rho^2}{3Q}-Ts-\mu\rho = -(Q+r_h)^3-\frac{r_h\beta^2}{2}\,.
\end{align}
As shown in the right panel of Fig. \ref{tQregion} the positive $\tQ$ solutions always correspond to the ground state, where $\delta \grandpot = \grandpot(\tQ>0) - \grandpot(\tQ<0)$. At zero density, there is no positive $\tQ$ for $\beta/4\pi T < 1/\sqrt{2}$. In this case $\tQ=0$ is the ground state.
%

%%%%%%%%%%%%%%%%%%%%%%%%%%
%%%%%%%%%%%%%%%%%%%%%%%%%%

To compute the diffusion constants $D_\pm$ we first need to compute thermodynamic susceptibilities \eqref{sus1} as a function of $T$ and $\mu$. Because the charge density $\rho$ \eqref{rho2} and entropy $s$ \eqref{s2} are functions of $r_h$ and $Q$ it will be convenient to express them as a function of $T$ and $\mu$.  In principle, $r_h$ and $Q$ can be expressed in terms of $T$ and $\mu$ from  \eqref{T2} and \eqref{mu2} but  their analytic expressions are very complicated except for $\mu=0$.  Therefore we will not present their expressions here and show some plots in the following subsections.  

From the general formula \eqref{gencond}, the electrical, thermal, and thermoelectric conductivities read
\begin{equation}
\begin{split}
	\sigma &= e^{\phi(r_h)}+\frac{4\pi\rho^2}{\beta^2 s}=\sqrt{1+\frac{Q}{r_h}}\left(1+\frac{\mu^2}{\beta^2}\right)\,, \\
	\overline \kappa &= \frac{4\pi s T}{\beta^2}\,,  \qquad \alpha = \frac{4\pi \rho}{\beta^2}\,. 
\end{split}
\end{equation}
and from \eqref{BV1} the butterfly velocity is
\begin{equation}
v_B^2=\frac{4\pi T}{Q+4r_h}\sqrt{\frac{r_h}{Q+r_h}}\,.
\end{equation}

\subsection{Zero density}
Let us first consider a neutral case, i.e. $\mu = \rho = 0$. As shown in the previous subsection and \eqref{mu2}, there are two solutions: $\tQ = 0$ and $\tQ = -1 + \frac{\tbeta}{\sqrt{2}}$.

For $\tQ=0$, the dilaton field $\phi$ vanishes and the solution \eqref{GReq} is reduced to (\ref{axion}) with $\mu=0$. This solution has  been considered in \cite{Blake:2016sud}. In this case, the transport coefficients and susceptibilities 
are given as
\begin{align}
\sigma &=1\,, \quad \kappa = \frac{4\pi^2T}{9\beta^2} \left(4\pi T+\sqrt{6\beta^2+16\pi^2 T^2} \right)^2 \,,\\
\chi &=\frac{1}{6} \left(4\pi T+\sqrt{6\beta^2+16\pi^2 T^2} \right) \,, \quad c_\rho =\frac{8\pi^2 T \left(4\pi T+\sqrt{6\beta^2+16\pi^2 T^2} \right)^2}{9\sqrt{6\beta^2+16\pi^2 T^2}} \,,
\end{align}
with $\alpha=0$ and $\zeta=0$. 
Thus the charge and energy diffusion constants are\footnote{The energy diffusion constant was also computed in \cite{Davison:2014lua}.}
\begin{equation}
\begin{split}
D_c &= \frac{\sigma}{\chi} = \frac{6}{4\pi T+\sqrt{6\beta^2+16\pi^2 T^2}} % \sim  \frac{1}{T} \frac{\sqrt{6}}{\beta/T} 
\,,\\
D_e &= \frac{\kappa}{c_\rho} = \frac{\sqrt{6\beta^2+16\pi^2 T^2}}{2\beta^2} % \sim  \frac{1}{2T} \frac{\sqrt{6}}{\beta/T} 
\,.
\end{split}
\end{equation}
%
%%%%%%%%%%%%%%%%%%%%%%%%%%
\begin{figure}
	\begin{center}
        {  \subfigure[ $\tQ=0$ ]
		{\includegraphics[width=6cm]{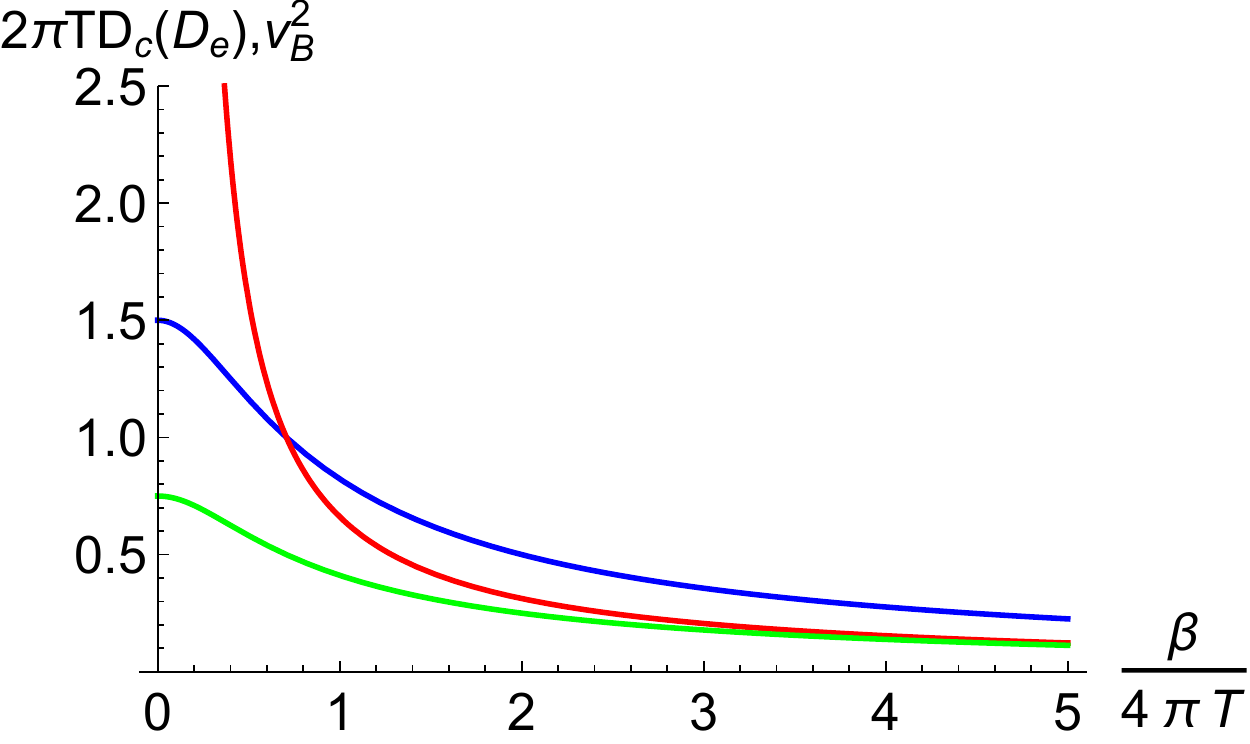}
                 \includegraphics[width=6cm]{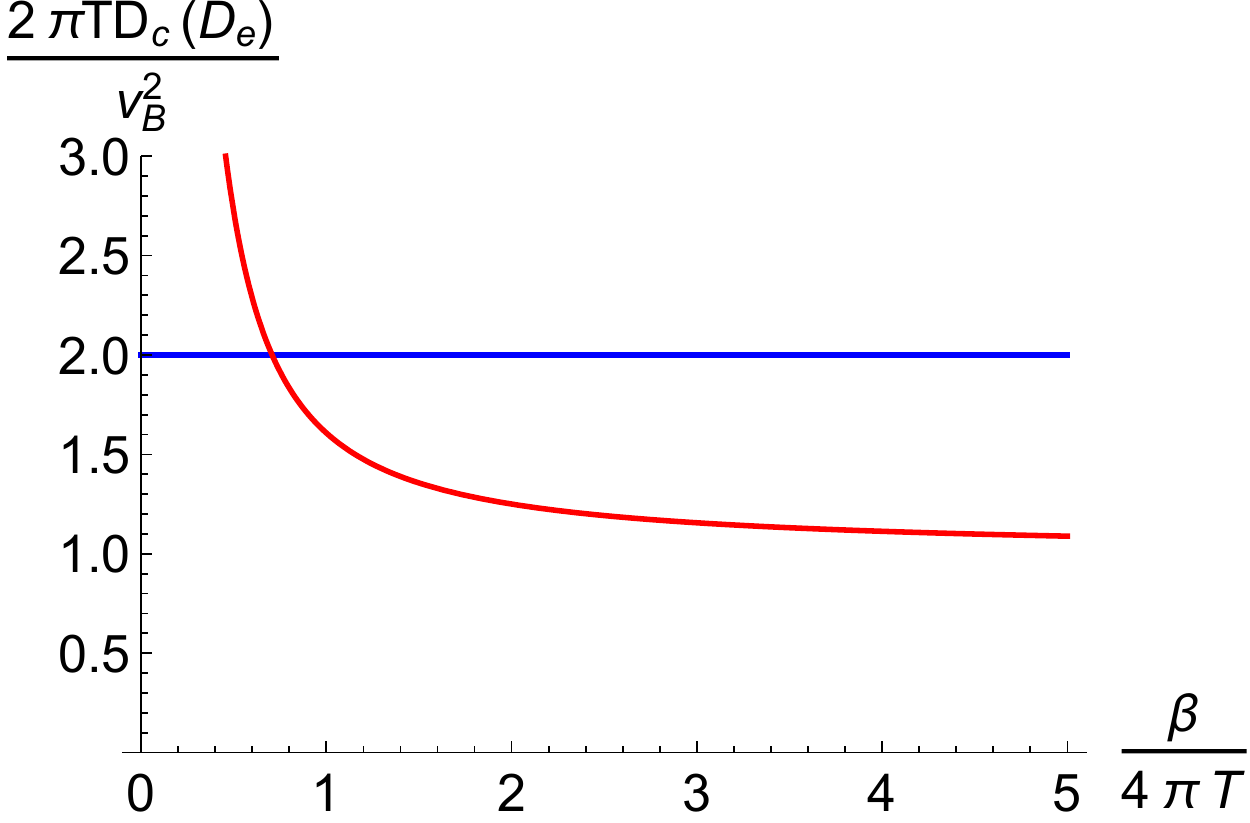}}
          \subfigure[ $\tQ = -1 + \frac{\tbeta}{\sqrt{2}}$ ]
		{\includegraphics[width=6cm]{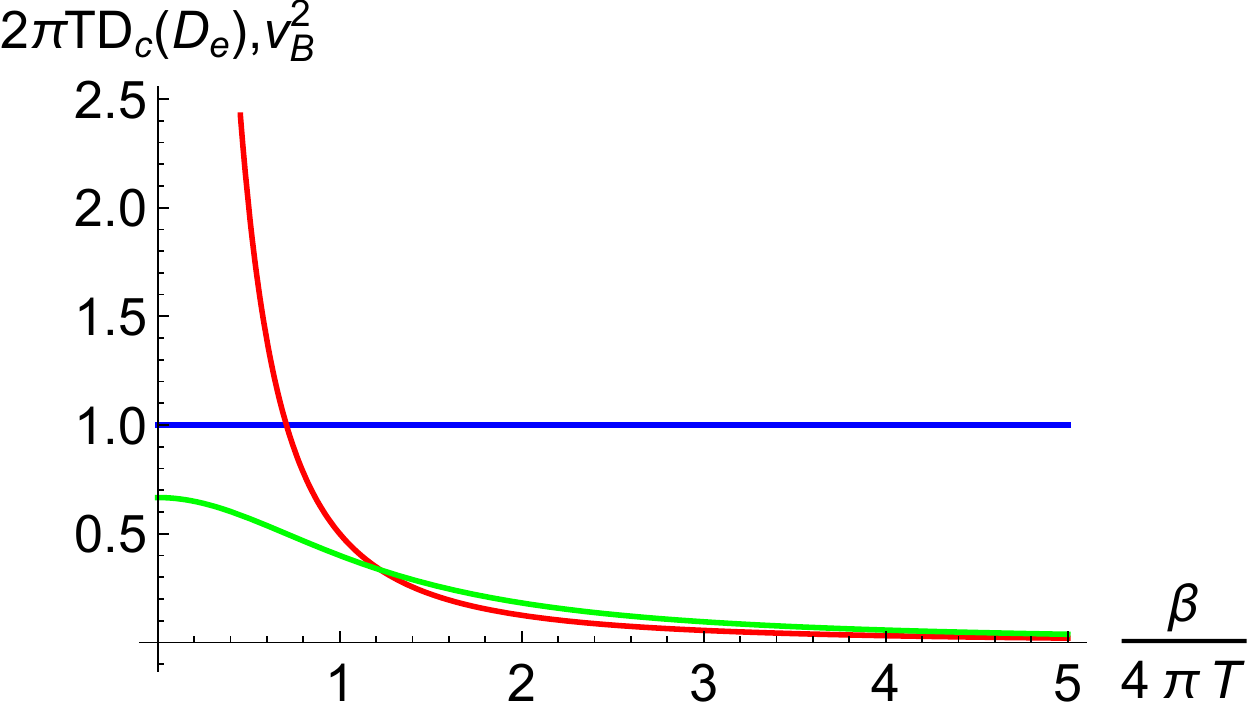}
		\includegraphics[width=6cm]{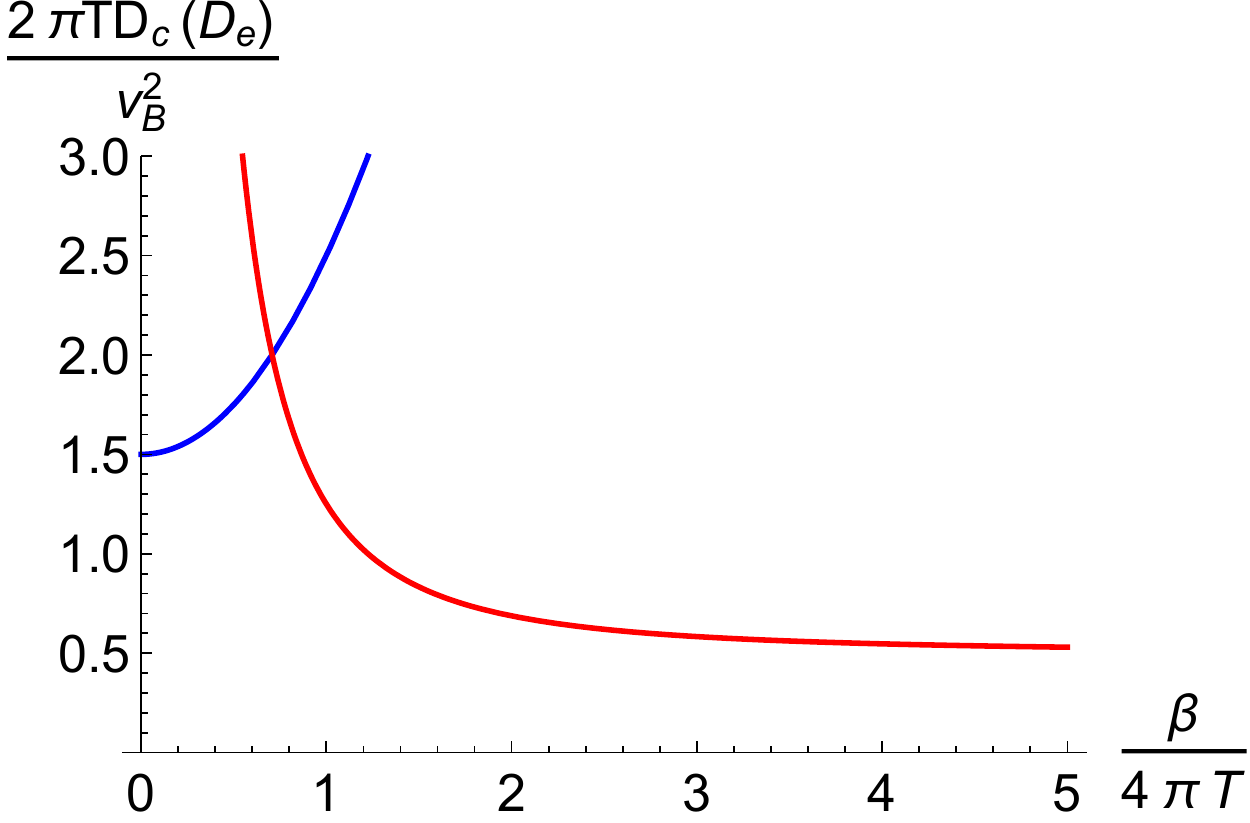}}
	\subfigure[ Ground state: there is a phase transition at $\beta/(4\pi T) = 1/\sqrt{2}$.]
		{\includegraphics[width=6cm]{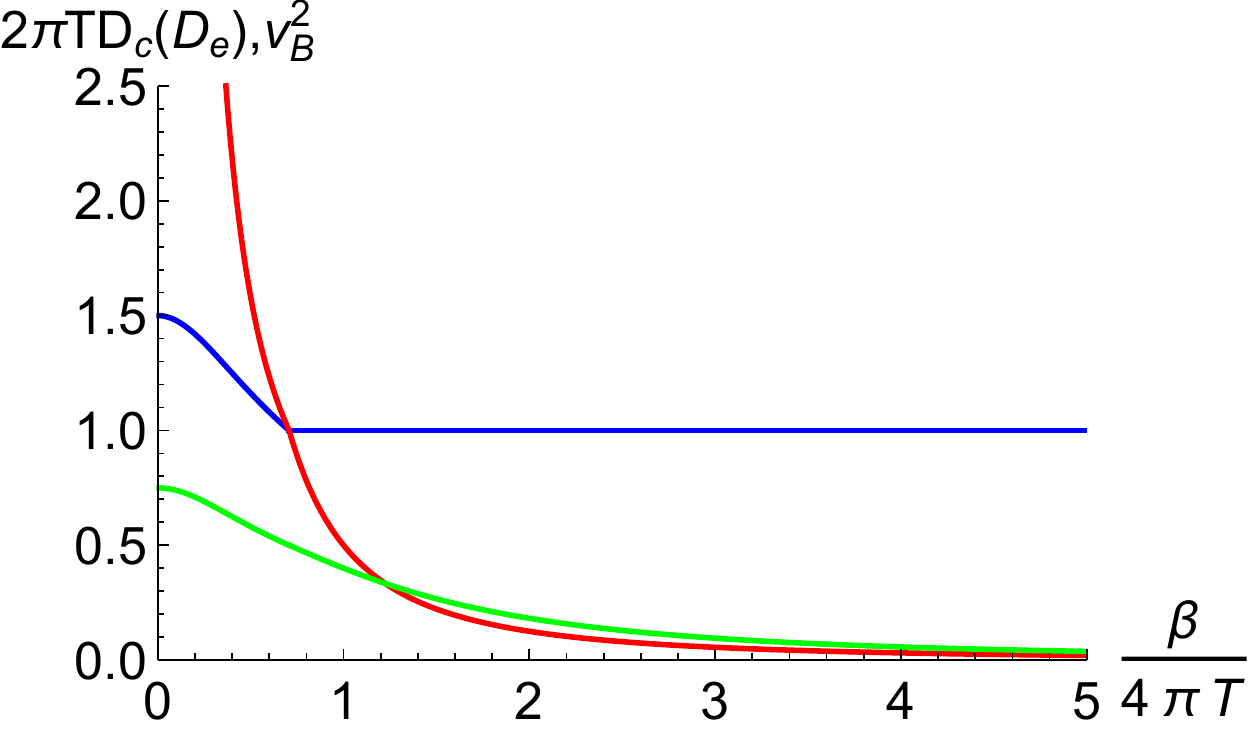}
		\includegraphics[width=6cm]{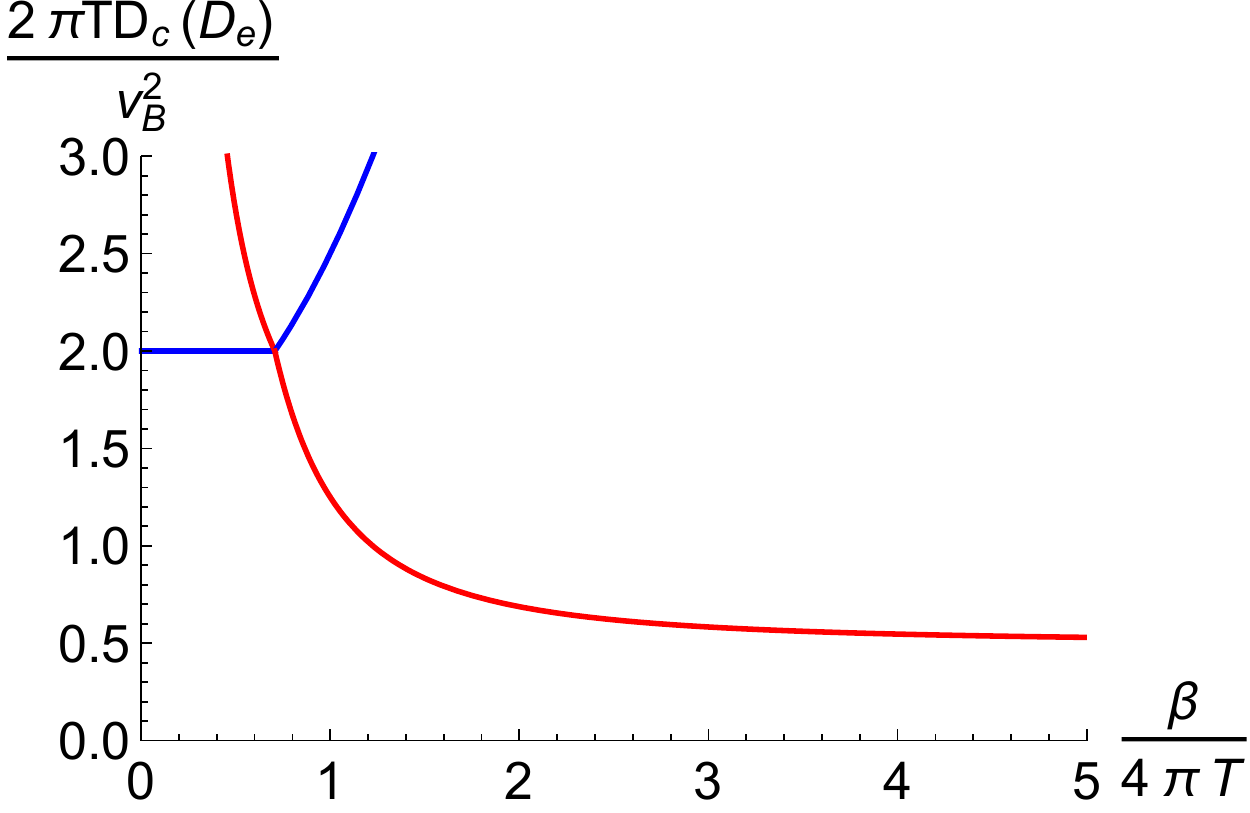}}}
	\end{center}
	\caption{Diffusion constants of the axion-dilaton model at zero density for $\tQ=0$ (a), for $\tQ = -1 + \frac{\tbeta}{\sqrt{2}}$ (b), and for the ground state (c). The blue curve is for $D_c$ and the red curve is  for $D_e$. The green curve displays $v_B^2$. } \label{fig4}
\end{figure} 
%%%%%%%%%%%%%%%%%%%%%%%%%%
%
The butterfly velocity is 
\begin{equation}
v_B^2 = \frac{6\pi T}{4\pi T+\sqrt{6\beta^2+16\pi^2 T^2}} %  \sim  \frac{\sqrt{6} \pi }{\beta/T}
\,,
\end{equation}
so 
\begin{equation}
\begin{split}
& \frac{2\pi T D_c}{v_B^2} = 2 \,, \\
& \frac{2\pi T D_e}{v_B^2} = 1+\frac{2\pi T}{3\beta^2}\left(4\pi T+\sqrt{6\beta^2+16\pi^2 T^2}\right)\,. 
\end{split}
\end{equation}
We show the plots for the diffusion constants and the butterfly velocity in Fig. \ref{fig4}(a).

For $\tQ = -1 + \frac{\tbeta}{\sqrt{2}}$, there is a nontrivial $\phi$ and $r_h =  2^{5/2} \pi^2 T^2 / \beta $.   The transport coefficients and susceptibilities are\footnote{The electric conductivity was also computed in \cite{Wu:2017mdl}}
\begin{align}
\sigma &= %2^{-1/4}\sqrt{\beta/ r_h}=
\frac{\beta}{2\sqrt{2}\pi T} \,, \qquad  \kappa = \frac{16\sqrt{2}\pi^3 T^2}{\beta}\,,  \\
\chi &=\beta/\sqrt{2}\,,  \qquad c_\rho = 4\sqrt{2}\pi^2 T \beta \,, 
\end{align}
with $\alpha=0$ and $\zeta=0$. Thus the charge and energy diffusion constants are\footnote{A similar result was obtained in \cite{Amoretti:2014ola}, where the graviton mass term was added instead of the linear axion term.}
\begin{equation}
\begin{split}
D_c &=\frac{\sigma}{\chi}=\frac{1}{2\pi T}\,,\\
D_e &=\frac{\kappa}{c_\rho}=\frac{4\pi T}{\beta^2}\,.
\end{split}
\end{equation}
The butterfly velocity is 
\begin{equation}
v_B^2 = \frac{16\pi^2 T^2}{24\pi^2 T^2+\beta^2}\  \rightarrow\ \frac{16 \pi^2 T^2}{\beta^2}\ \mathrm{for}\ \beta/T \gg 1 
 \,,
\end{equation}
so
\begin{equation} \label{zerod1}
\begin{split}
& \frac{2\pi T D_c}{v_B^2} = \frac{\beta^2}{16\pi^2T^2}    + \frac{3}{2}\  \rightarrow\  \frac{\beta^2}{16\pi^2T^2}  \ \mathrm{for}\ \beta/T \gg 1 
 \,,  \\
& \frac{2\pi T D_e}{v_B^2} = \frac{1}{2}+\frac{12\pi^2T^2}{\beta^2}   \  \rightarrow\  \frac{1}{2} \ \mathrm{for}\ \beta/T \gg 1 
\,.
\end{split}
\end{equation}
We show the plots for the diffusion constants and the butterfly velocity in Fig. \ref{fig4}(b).

By comparing the grand potential of two cases, $\tQ=0$  and $\tQ = -1 + \tbeta/\sqrt{2}$, we find that the $\tQ=0$ case
is the ground state for $\beta/(4\pi T) < 1/\sqrt{2}$ and  the $\tQ = -1 + \tbeta/\sqrt{2}$ case  is the ground state for $\beta/(4\pi T) > 1/\sqrt{2}$, similarly to Fig. \ref{tQregion}(a). Taking this phase transition into account we show the the final results in Fig. \ref{fig4}(c). 

Let us summarize the asymptotic behavior in the incoherent regime $\beta/T \gg 1$. For $\tQ=0$, $D_c, D_e \sim T/\beta$ and $v_B^2 \sim T/\beta$, so both $D_c/v_B^2$ and $D_e/v_B^2$ saturate their bounds.
For $\tQ = -1 + \tbeta/\sqrt{2}$, $D_c \sim 1/T$ and  $D_e \sim T/\beta^2$ while $v_B^2 \sim T^2/\beta^2$, so only  $D_e/v_B^2$ saturates its bound. However, in the incoherent regime, $\tQ = -1 + \tbeta/\sqrt{2}$ case is the stable state.  Note that, in the incoherent regime, the charge diffusion ($2\pi T D_c$) saturates the bound but $2\pi T D_c/v_B^2$ diverges, while $2\pi T D_e/v_B^2$ saturates the bound.

%%%%%%%%%%%%%%%%%%%%%%%%%%
%\begin{figure}\label{}
%	\begin{center}
%		\includegraphics[width=4.95cm]{gubnfe.pdf}
%		\includegraphics[width=4.95cm]{gubn3.pdf}
%		\includegraphics[width=4.95cm]{gubn4.pdf}
%	\end{center}
%	\caption{Gibbs free energy density and diffusion constants of neutral Axion-Dilaton model.}
%\end{figure} 
%%%%%%%%%%%%%%%%%%%%%%%%%%

\subsection{Finite density}

\begin{figure}	
	\begin{center}
		      {\subfigure[ Diffusion constants ($2\pi T D_{\pm}$) and the butterfly velocity ($v_B^2$) ]
			{\includegraphics[width=5cm]{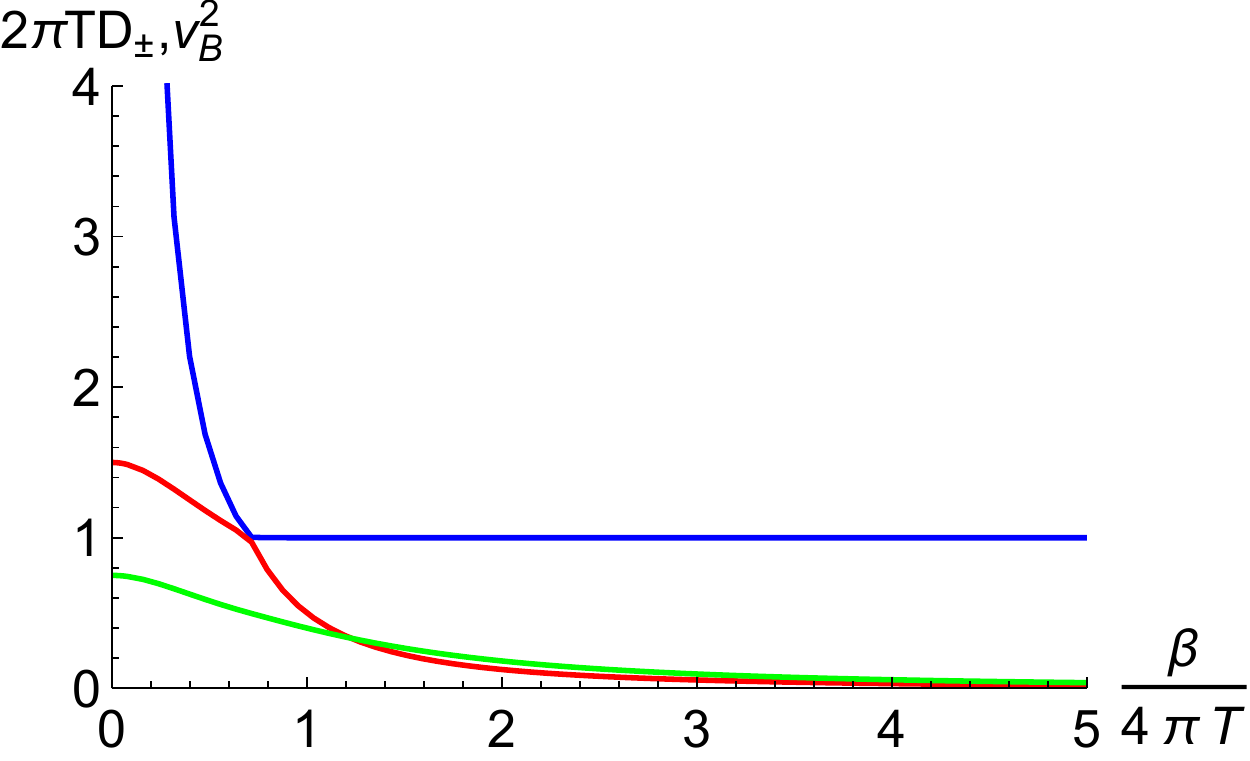}
			\includegraphics[width=5cm]{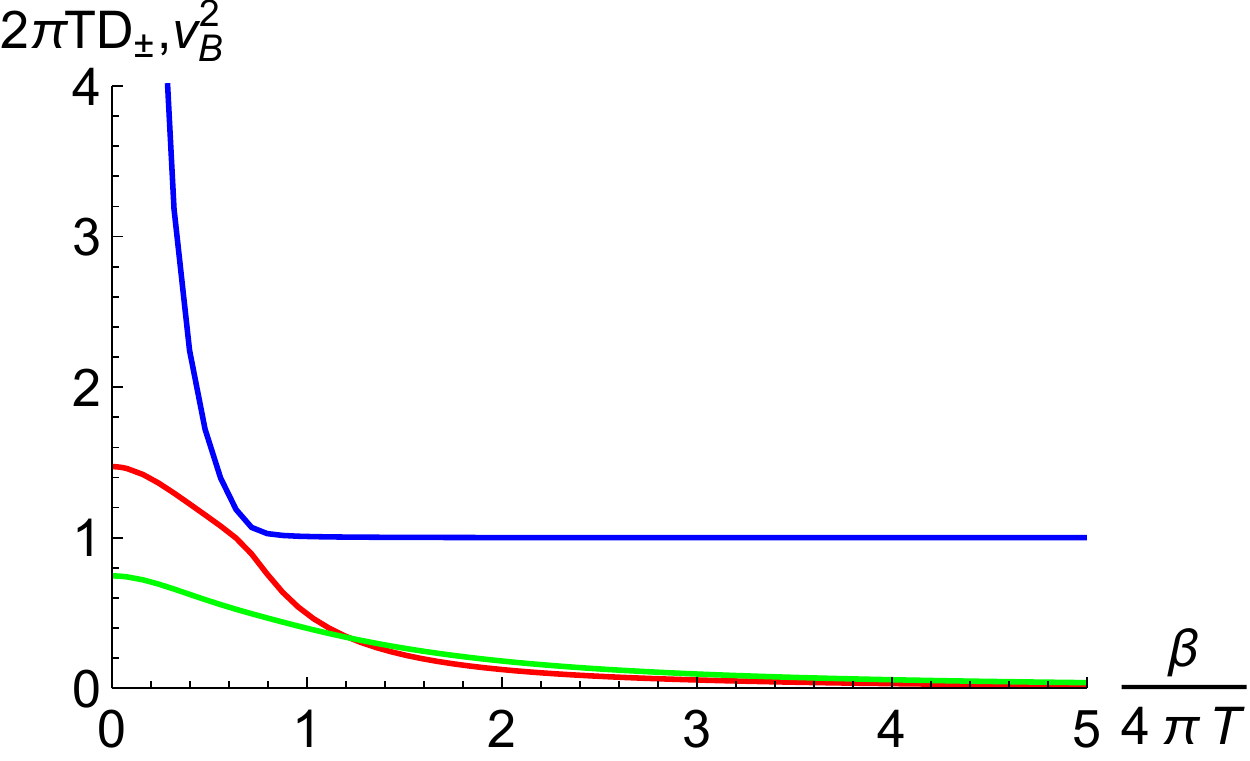}
			\includegraphics[width=5cm]{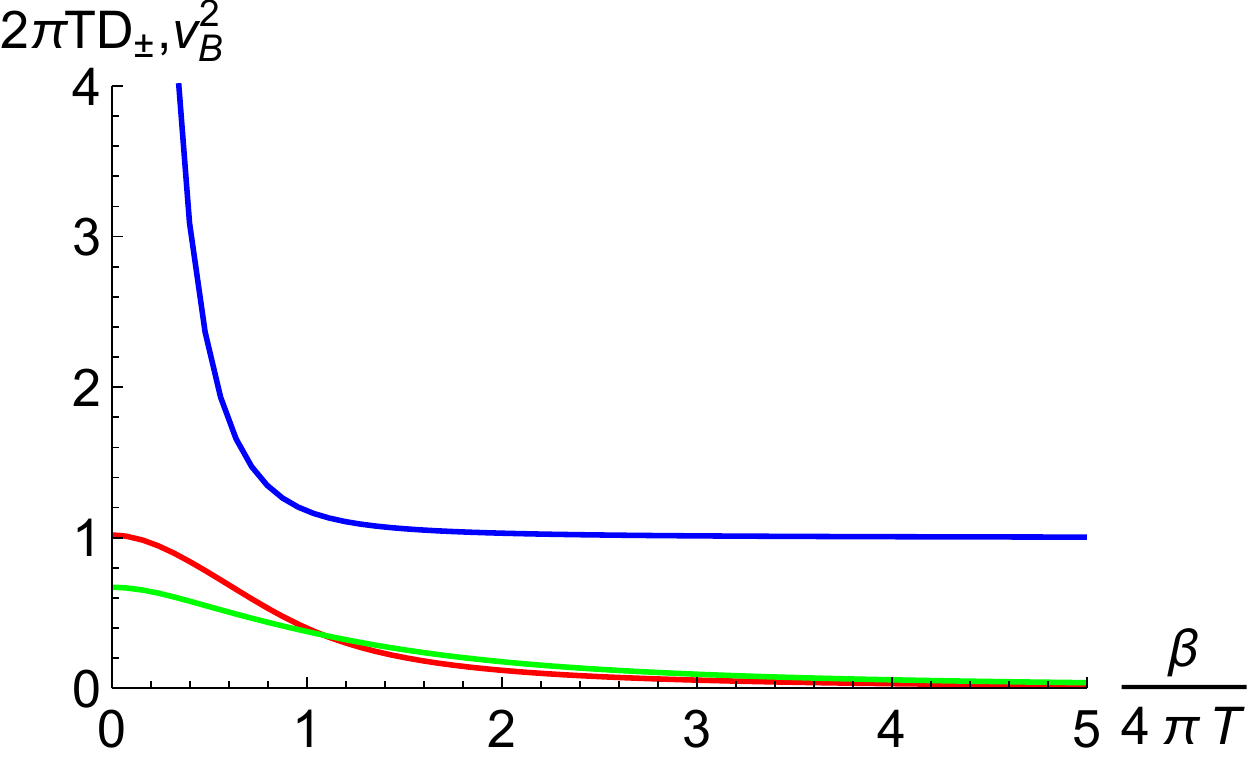}
			}
	              \subfigure[ Diffusion constants/(butterfly velocity)$^2$: $2\pi T D_c/v_B^2$ and $2 \pi T D_e/v_B^2$  ]
			{\includegraphics[width=5cm]{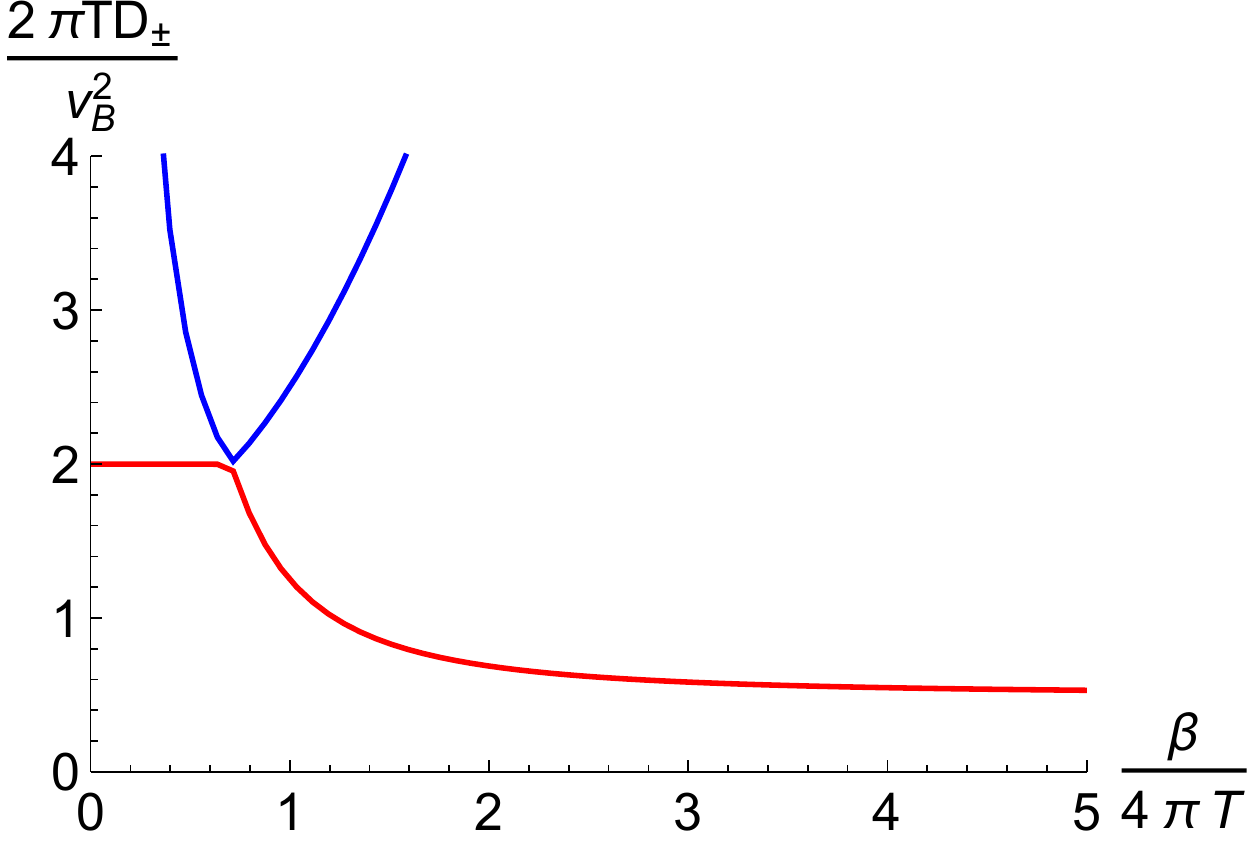}
			\includegraphics[width=5cm]{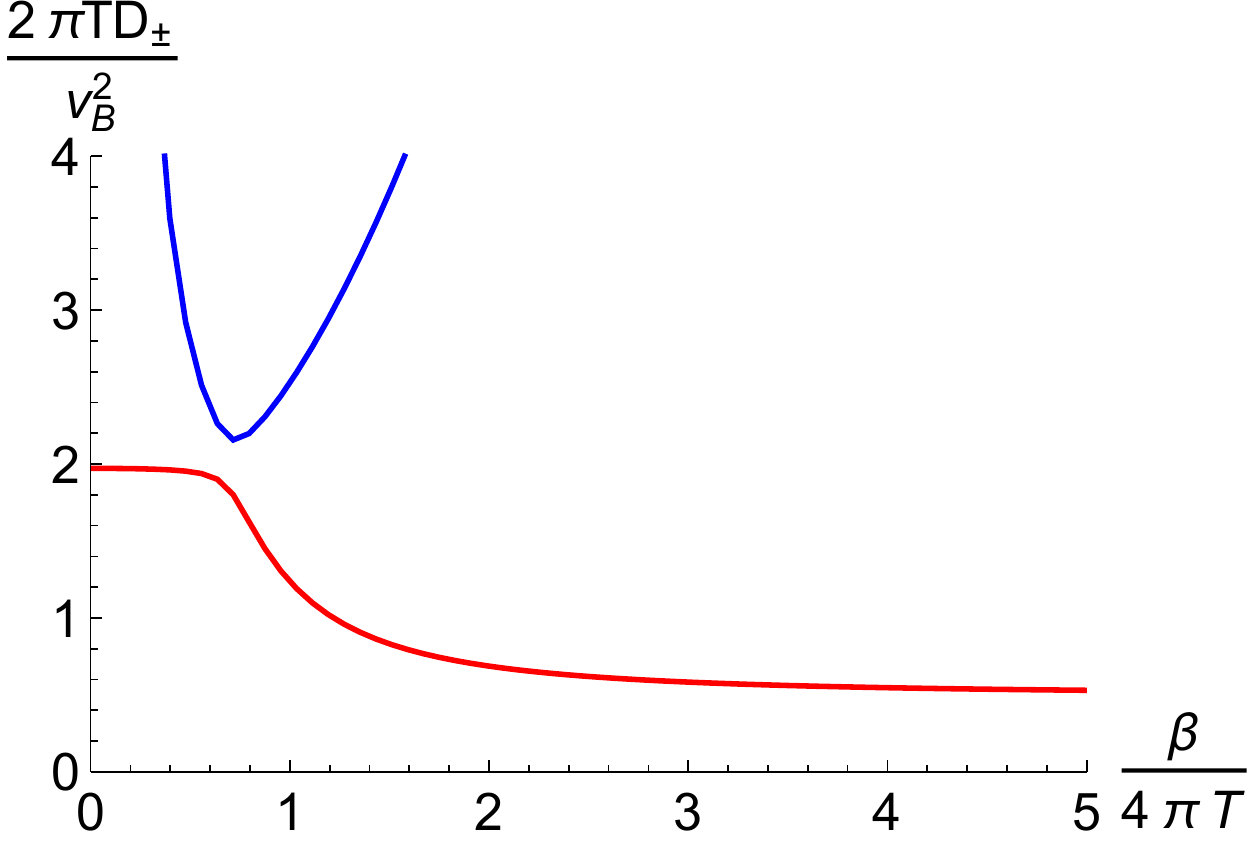}
			\includegraphics[width=5cm]{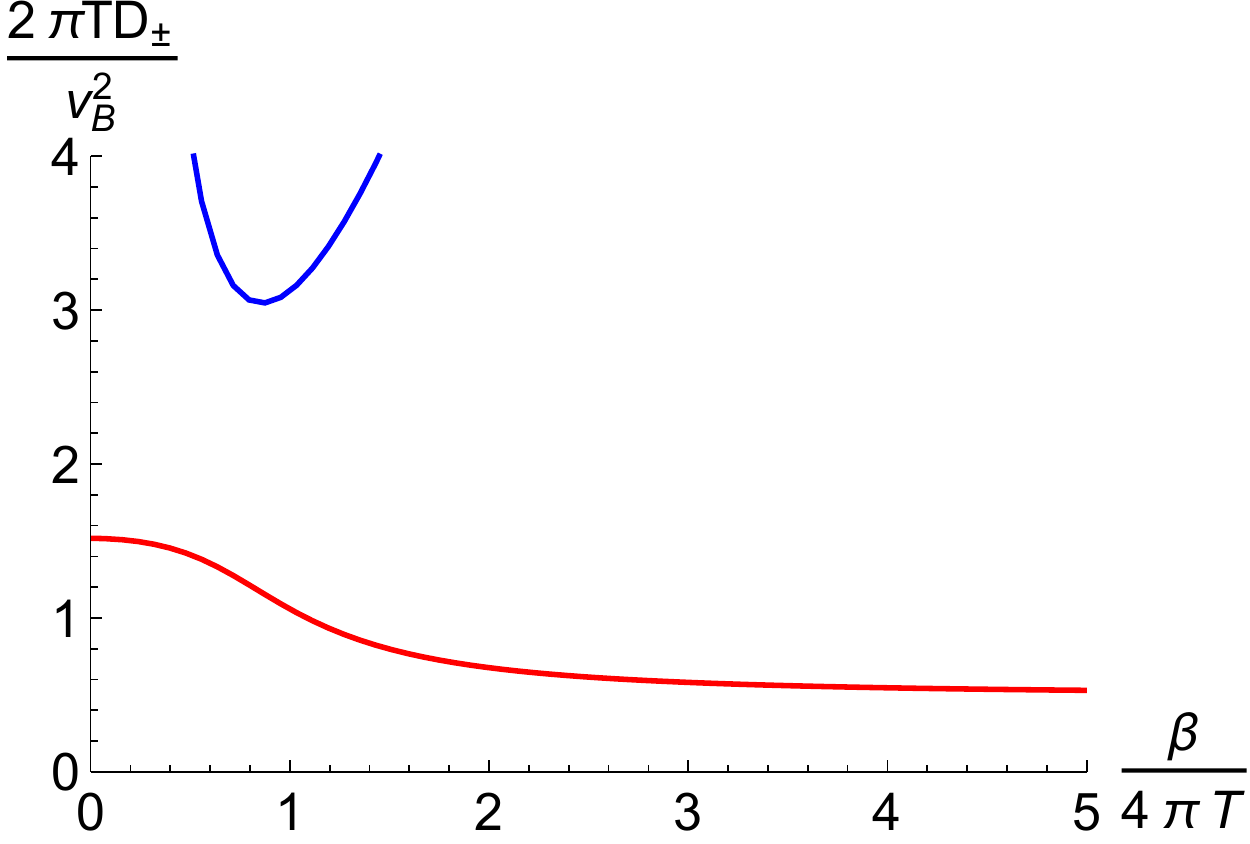}}}
	\end{center}
	\caption{Diffusion constants of the axion-dilaton model at finite density. $\mu/T=0.1, 1, 5$ from left to right. The blue curve is for $D_+$ and the red curve is for $D_-$. The green curve displays $v_B^2$.  $2\pi D_+$ and $2\pi T D_{-}/v_B^2$ saturates the universal value in the incoherent regime:  $\beta/T \gg 1$ and $\beta/\mu \gg 1$.  } 	\label{fig5}
\end{figure}

The horizon position $r_h$ can be expressed in terms of $T$ and $\mu$ by eliminating $Q$ in \eqref{T2} and \eqref{mu2}. 
However,  unlike the previous cases, if $\mu \ne 0 $ the expression of $r_h$ is long and complicated. Furthermore, to compute thermodynamic susceptibilities we need to perform differentiations where a few steps of implicit differentiations and chain rules are involved.  As a result, all the final results are analytic but they are not so illuminating.  Therefore, instead of presenting analytic expressions we show the plots in Fig. \ref{fig5}.  
At finite density ($\mu\ne0$) there are two branches, positive $Q$ and negative $Q$, as explained in \eqref{branches} and Fig. \ref{tQregion}.  Because the positive $Q$ branch is the stable solution we showed it in Fig. \ref{fig5}.

Even though the exact analytic formulas are complicated their large $\beta$ limit, which we are interested in for the universal bound, can be read off analytically.  For large $\beta$ in the positive $Q$ branch, $r_h/\beta \rightarrow 0$ and $Q/\beta \rightarrow  1/\sqrt{2}$ from \eqref{T2} and \eqref{mu2}, which yield
\begin{equation}
\begin{split}
\sigma &\sim \frac{\beta}{2\sqrt{2}\pi T} \,, \qquad  \kappa \sim \frac{16\sqrt{2}\pi^3 T^2}{\beta}\qquad  \alpha \sim \frac{4\pi\mu}{\sqrt{2}\beta} \,,  \\
\chi &\sim \beta/\sqrt{2}\,,  \qquad c_\rho \sim 4\sqrt{2}\pi^2 T \beta \,, \qquad  \zeta \sim \frac{64\sqrt{2}\pi^4T^3\mu}{\beta^3} \,.
\end{split}
\end{equation}
Therefore, the charge and energy diffusion constants are
\begin{equation}
D_+  \sim \frac{1}{2\pi T}\,, \qquad D_-  \sim \frac{4\pi T}{\beta^2}\,.
\end{equation}
Together with the butterfly velocity at large $\beta \gg \mu$  
\begin{equation}
v_B^2  \sim \frac{16\pi^2 T^2}{\beta^2} \,,
\end{equation}
we have
\begin{equation}
\frac{2\pi T D_+}{v_B^2} \sim  \frac{\beta^2}{16\pi^2T^2} \,, \qquad  \frac{2\pi T D_e}{v_B^2} \sim \frac{1}{2}+\frac{12\pi^2T^2}{\beta^2} 
\,.
\end{equation}
Thus we find that the bounds for $D_\pm$ at zero density \eqref{zerod1} still hold at finite density. The correction by $\mu$ is at higher order and the results are robust in the incoherent regime. 

\begin{figure}
	\begin{center}
		    {  \subfigure[ Diffusion constants: $2\pi T D_c$ and $2\pi T D_e$ ]
			{\includegraphics[width=5cm]{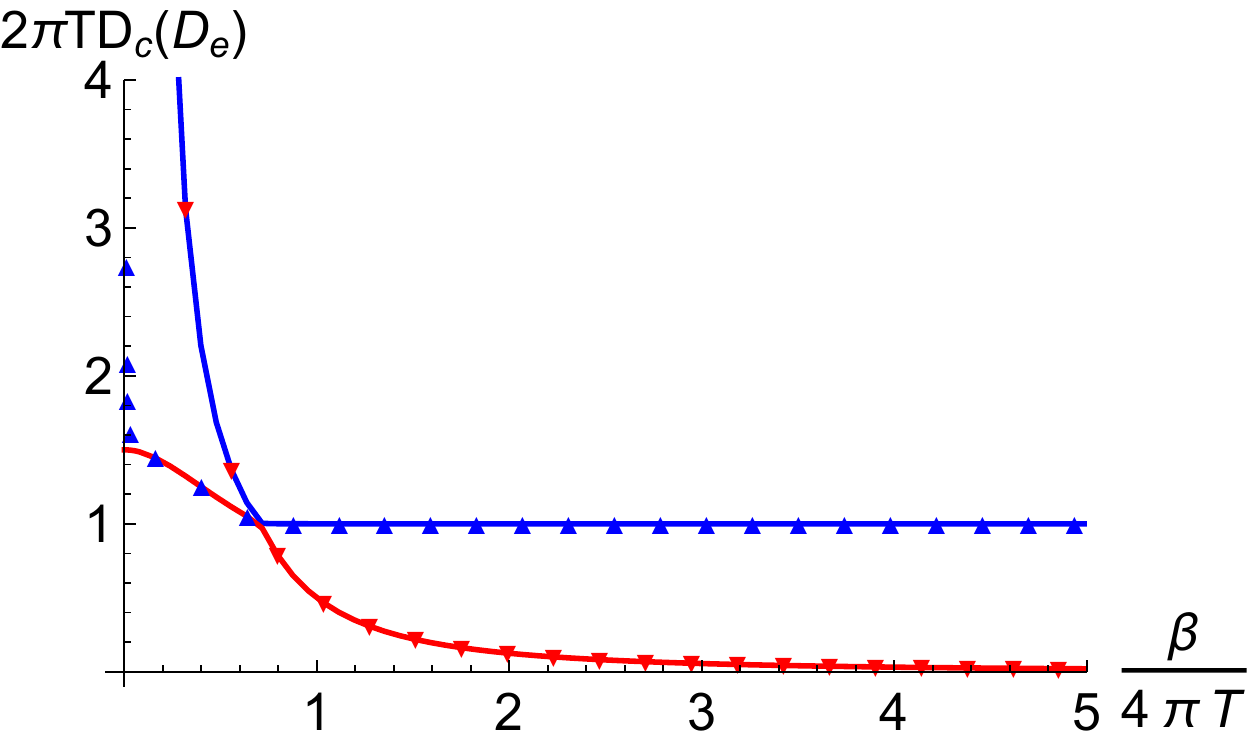}
			\includegraphics[width=5cm]{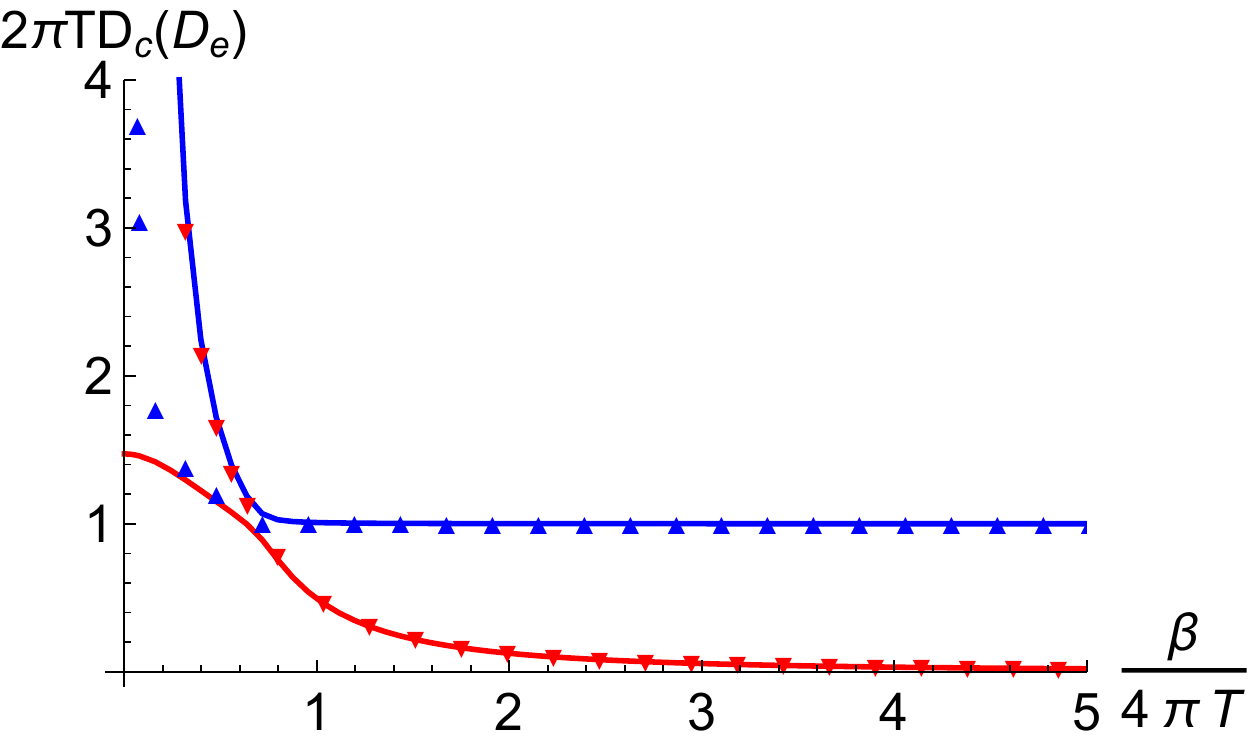}
			\includegraphics[width=5cm]{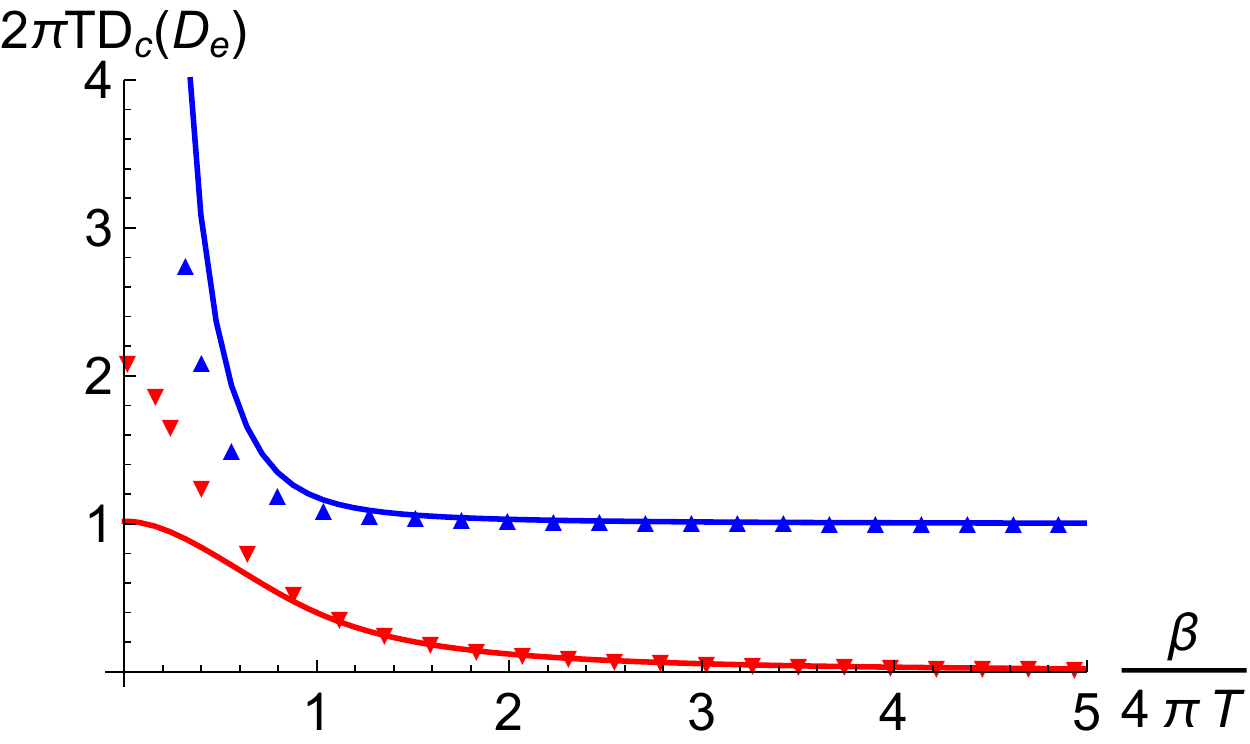}
			}
	              \subfigure[  Diffusion constants/(butterfly velocity)$^2$: $2\pi T D_c/v_B^2$ and $2 \pi T D_e/v_B^2$  ]
			{\includegraphics[width=5cm]{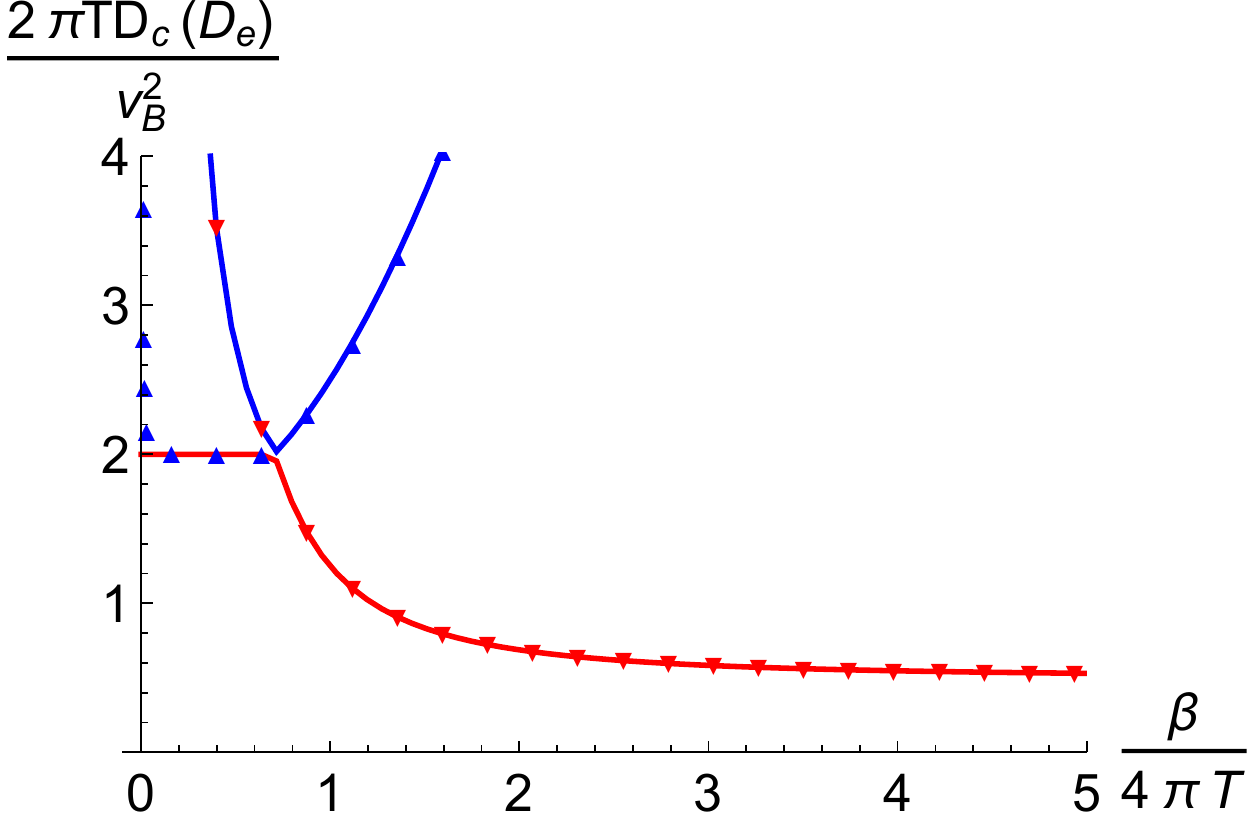}
			\includegraphics[width=5cm]{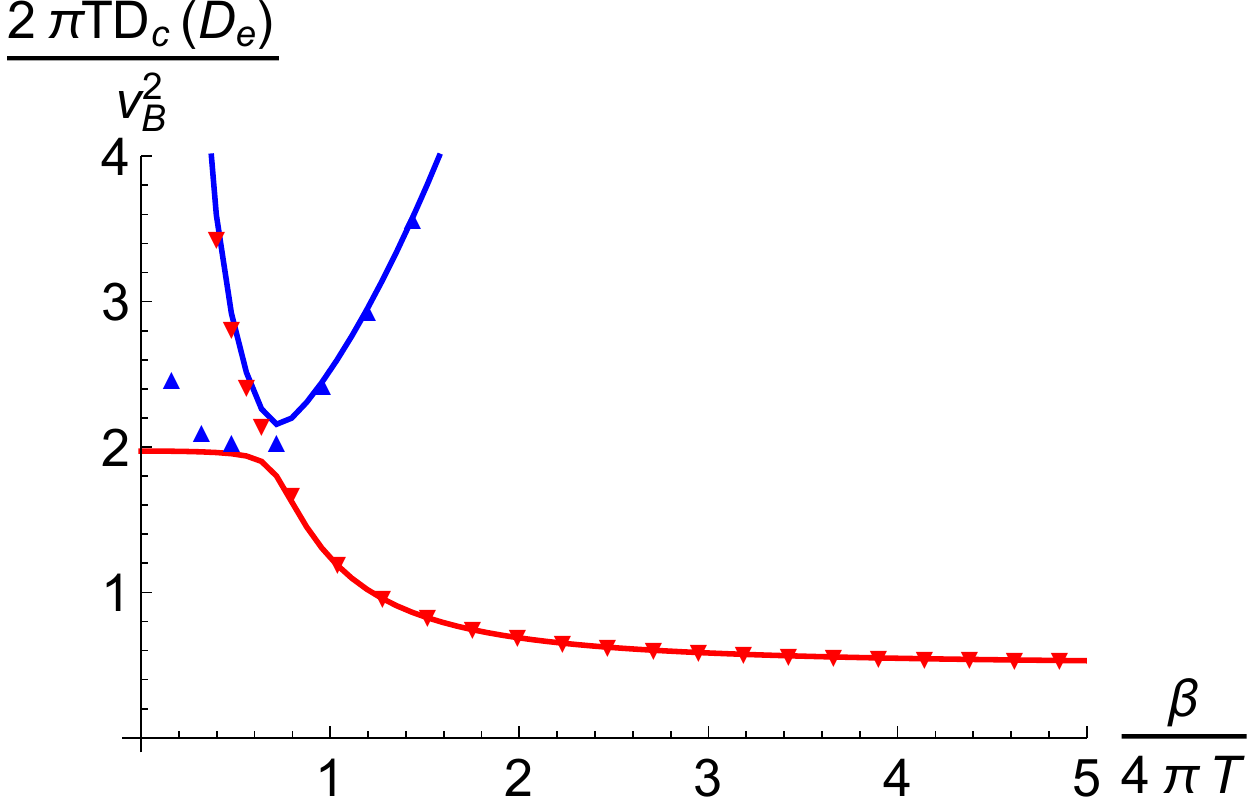}
			\includegraphics[width=5cm]{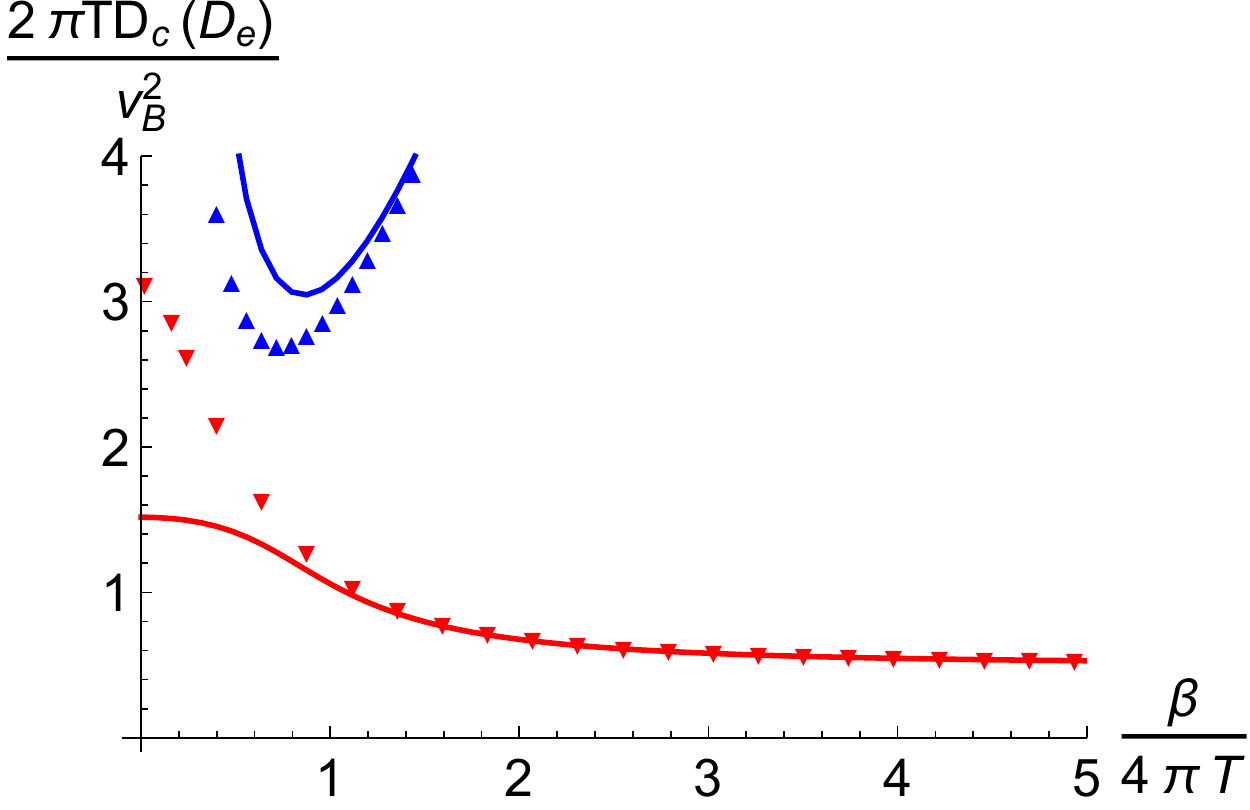}}}
	\end{center}
	\caption{Diffusion constants of the axion-dilaton model with/without a mixing term. $\mu/T=0.1, 1, 5$ from left to right. The blue curve is for $D_c$ and the red curve is for $D_e$. The triangles display the results without the mixing term $\mathcal{M}$ in  \eqref{c1c22}. For comparison, we also display  the results with the mixing term, the solid curves in Fig. \ref{fig5}. } 	\label{fig6}
\end{figure} 

Like the linear axion model, in the incoherent regime, the mixing term between the charge and energy diffusion, $\mathcal{M}$ in \eqref{c1c22}, is negligible so $D_+$ and $D_-$ may be identified with $D_c$ and $D_e$ respectively.  To see the effect of the mixing term, we show another plot for $D_c$ and $D_e$ in Fig. \ref{fig6}. Like Fig. \ref{fig2}, the triangles are the results without the mixing term, where we simply identify $D_c \equiv \sigma/\chi$ and $D_c \equiv \kappa/c_\rho$. Like the linear axion model, below $\beta/4\pi T \lesssim 1 $ the mixing term is important and this mixing effect decreases as $\mu/T$ increases. Note that, for small $\mu/T=0.1$, this mixing effect is `maximal' in the sense $D_-$ is identified with $D_c$ and $D_+$ is identified with $D_e$.

In this model, the relation between the linear-$T$-resistivity and the universality of diffusion \eqref{nice1} is not realized. Even though $\sigma \sim 1/T$, $C_+$ is not bounded because $\chi v_B^2 \sim T^2/\beta^2$. It is possible that there is a more relevant velocity scale than the butterfly velocity for the charge diffusion.  If that velocity scale is independent of temperature, \eqref{nice1} can be realized because $\chi$ is temperature-independent in this model.

\section{Conclusion} \label{conc}

In this paper, we have studied two diffusion constants and the butterfly velocity  at {\it finite density} in two holographic models: linear axion model and axion-dilaton model. In both cases, the axion field is of the form $\psi_I =  \beta \delta_{Ii} x^i$ and plays a role of momentum relaxation, where large $\beta$ means large momentum relaxation. 
At zero density, the axion-dilaton model undergoes a phase transition to the linear axion model when momentum relaxation is weak. At finite density, the axion-dilaton model has two branches of solutions so we have to choose a ground state by comparing their grand potentials. 

There are two diffusion constants $D_\pm$ describing the coupled diffusion of charge and energy.
We have showed the exact relation between $D_\pm$ and $(D_c, D_e)$ in Fig. \ref{fig2} and Fig. \ref{fig6}. 
In the incoherent regime, the mixing between  charge  and thermal diffusion is suppressed so $D_+$ and $D_-$ can be identified with $D_c$ and $D_e$ respectively. However, in the coherent regime the effect of the mixing term becomes strong so $D_c$ and $D_e$ can not be decoupled and $D_\pm$ should be considered. 
In particular, at very small $\mu/T$, this mixing effect is `maximal' in the sense that $D_-$ is identified with $D_c$ and $D_+$ is identified with $D_e$, which is opposite to the case in the incoherent regime.  
In Table \ref{table1}  the diffusion constants and the butterfly velocity of two models at finite density in the incoherent regime are summarized.

\begin{table}[]
\begin{center}
\begin{tabular}{|c||c|c|c|c|c|}
\hline
        & $D_+(\approx D_c)$ & $D_-(\approx D_e)$ & $v_B$ &   $C_+ \left(\frac{2\pi T D_+}{v_B^2}\right)$   &  $C_-\left(\frac{2\pi T D_-}{v_B^2} \right) $\\ 
 \hline
 \hline
 Linear axion & $\frac{\sqrt{6}}{\beta }$ &  $\frac{\sqrt{3/2}}{\beta }$ & $ \sqrt[4]{6}\sqrt{\pi} \sqrt{\frac{ T}{\beta }}$ & 2  &  1  \\ 
 \hline
 Axion-dilaton & $\frac{1}{2\pi T}$ &  $\frac{4\pi T}{\beta^2}$ &  $\frac{4\pi T}{\beta}$  &  $\frac{\beta^2}{16\pi^2T^2}$ 
  &  $\frac{1}{2}$ \\ 
 \hline
\end{tabular}
\end{center}
\caption{Summary of the results: diffusion constants for the linear axion model and axon-dilaton model in the incoherent regime $\beta/T \gg 1$ and $\beta/\mu \gg 1$. }
\label{table1}
\end{table}%
The thermal diffusion constant   at finite density in the incoherent regime can be written as
\begin{equation}
D_- \approx D_e \approx C_- \frac{v_B^2}{2\pi T} \,,
\end{equation}
where $C_- = 1$ for the linear axion model and $C_-=1/2$ for the axion-dilaton model. These agree to the values at zero density and $C_-$ is universal independently of density.
In \cite{Blake:2016jnn}, it was shown that the holographic models with  IR geometry of AdS$_2 \times R^d$, which can be supported by finite density and/or axion field, give $1/2< C_- \leqslant 1$ at low temperature limit. 
Because the IR geometry of the axion model is AdS$_2 \times R^d$, $C_-=1$ can be anticipated from \cite{Blake:2016jnn}, where the low temperature limit of the linear axion model was considered. 
Here, we analysed the model at {\it any temperature}  for both  the coherent and the incoherent regime. The IR geometry of the axion-dilaton model in this paper is conformal to  AdS$_2 \times R^d$ so the model does not belong to the class (the models with  IR geometry of AdS$_2 \times R^d$) studied in \cite{Blake:2016jnn}. However, our result $C_-=1/2$ yields the lower bound of the range that the class allows i.e. $1/2< C_- \leqslant 1$.

The charge diffusion constant is written as
\begin{equation}
D_+ \approx D_c \approx C_+ \frac{v_B^2}{2\pi T} \,,
\end{equation}
where $C_+ = 2$ for the linear axion model and $C_+= \beta^2/16\pi^2 T^2$ for the axion-dilaton model. These agree to the results at zero density. $C_+$ of the axion model is universal at any finite density but $C_+$ in the axion-dilaton model increases as $\beta/T$ increases. However, the charge diffusion constant of the axion-dilaton model itself saturates to the universal lower bound:
\begin{equation}
D_+ \approx D_c \approx  \frac{1}{2\pi T} \,.
\end{equation}
Thus, if there is more relevant velocity scale than the butterfly velocity and if it is temperature-independent, the relation between the universality of charge diffusion and linear-$T$-resistivity \eqref{nice1} may be realized in this model ($\chi$ is independent of  temperature in this model). Indeed, it may be possible that there is another velocity scale for charge diffusion by the following observations. 

To understand the universality of energy diffusion, it is important to note that two susceptibilities, $\chi$ and $\zeta$, cannot be written in terms of the horizon data while $c_\mu$ can be written only in terms of the horizon data as shown in \eqref{sus1}.  To compute $\chi$ and $\zeta$, the full bulk solution is needed. On the other hand, all conductivities and the butterfly velocity are written in terms of the horizon data as shown in \eqref{gencond} and \eqref{BV1}.
Because the energy diffusion constant {\it in the incoherent regime} is the ratio of $\kappa$ to $c_\rho \approx c_\mu$, it can be written only in terms of horizon data, while the charge diffusion constant can not.  Therefore, it is plausible that the energy diffusion constant in the incoherent regime may be universal thanks to a universal property of the black hole horizon.  In the coherent regime, the mixing term $\mathcal{M}$ \eqref{DDD1} is important and $(\zeta, \chi)$ should be taken into account to compute $c_\rho$ so the energy diffusion constant in the coherent regime will not be universal. By the same reason  the charge diffusion constant may not be universal and this non-universality was also observed in \cite{Lucas:2016yfl,Davison:2016ngz,Baggioli:2016pia} and it seems that chaos is only  connected to energy diffusion~\cite{Patel:2016aa,Davison:2016ngz}. Therefore, it is an interesting future direction to search 
a velocity scale for  charge diffusion. For energy diffusion, the bound based on the butterfly velocity seems to be more robust than charge diffusion.  However, recently a counter example was found in \cite{Gu:2017ohj} and it will be also interesting to understand the extent to which the energy diffusion bound with the butterfly velocity is robust~\cite{WIP}. 
 
%\newpage

\acknowledgments

We would like to thank Mike Blake, Blaise Gouteraux, and Wei-Jia Li for valuable discussions. The work was supported by Basic Science Research Program through the National Research Foundation of Korea(NRF) funded by the Ministry of Science, ICT $\&$ Future Planning(NRF- 2014R1A1A1003220) and GIST Research Institute(GRI) grant funded by the GIST in 2017.

\bibliographystyle{JHEP}
%\bibliography{/Users/FortOeFP/Dropbox/Research/Template/KyKimRefs}
%\bibliography{KyKimRefs_1}

\providecommand{\href}[2]{#2}\begingroup\raggedright\endgroup

\end{document}